\documentclass[fleqn,usenatbib]{mnras}
\usepackage[T1]{fontenc}
\usepackage{ae,aecompl}
\usepackage{multirow}
\usepackage{graphicx}	
\usepackage{amsmath}	
\usepackage{amssymb}	
\newcommand{\bz}{$\langle B_z \rangle$}

\newcommand{\Le}{$L_\mathrm{\textsc{ecme}}$}
\usepackage{booktabs,caption}
\usepackage[flushleft]{threeparttable}
\usepackage{newtxtext,newtxmath}

\title[Radio survey of hot magnetic stars]{Testing a scaling relation between coherent radio emission and physical parameters of hot magnetic stars}

\author[B. Das et al.]{
Barnali Das,$^{1}$\thanks{E-mail: barnali@udel.edu}
Poonam Chandra,$^{2,3}$
Matt E. Shultz,$^{1}$
Paolo Leto,$^4$
Zden\v ek Mikul\'a\v sek,$^{5}$
\newauthor
V\'eronique Petit$^{1}$ and
Gregg A. Wade$^{6}$
\\
$^{1}$Department of Physics and Astronomy, Bartol Research Institute, University of Delaware, 217 Sharp Lab, Newark, DE 19716, USA\\
$^{2}$National Centre for Radio Astrophysics, Tata Institute of Fundamental Research,  Pune University Campus, Pune-411007, India\\
$^{3}$National Radio Astronomy Observatory, 520 Edgemont Rd, Charlottesville, VA 22903, USA\\
$^{4}$INAF$-$Osservatorio Astrofisico di Catania, Via S. Sofia 78, I-95123 Catania, Italy\\
$^{5}$Department of Theoretical Physics and Astrophysics, Masaryk University, Kotl\'a\v rsk\'a 2, cz616 00 Brno, Czech Republic\\
$^{6}$Department of Physics and Space Science, Royal Military College of Canada, PO Box 17000, Station Forces, Kingston, ON K7K 7B4, Canada\\
}

\date{Accepted XXX. Received YYY; in original form ZZZ}

\pubyear{2022}

\begin{document}
\label{firstpage}
\pagerange{\pageref{firstpage}--\pageref{lastpage}}
\maketitle

\begin{abstract}
Coherent radio emission via electron cyclotron maser emission (ECME) from hot magnetic stars was discovered more than two decades ago, but the physical conditions that make the generation of ECME favourable remain uncertain.
Only recently was an empirical relation, connecting ECME luminosity with the stellar magnetic field and temperature, proposed to explain what makes a hot magnetic star capable of producing ECME. 
This relation was, however, obtained with just fourteen stars. Therefore, it is important to examine whether this relation is robust.
With the aim of testing the robustness, we conducted radio observations of five hot magnetic stars.
This led to the discovery of three more stars producing ECME. 
We find that the proposed scaling relation remains valid after the addition of the newly discovered stars.
However we discovered that the magnetic field and effective temperature correlate for $T_\mathrm{eff}\lesssim 16$ kK (likely an artifact of the small sample size), rendering the proposed connection between ECME luminosity and $T_\mathrm{eff}$ unreliable. By examining the empirical relation in light of the scaling law for incoherent radio emission, we arrive at the conclusion that both types of emission are powered by the same magnetospheric phenomenon. Like the incoherent emission, coherent radio emission is indifferent to $T_\mathrm{eff}$ for late-B and A-type stars, but $T_\mathrm{eff}$ appears to become important for early-B type stars, possibly due to higher absorption, or, higher plasma density at the emission sites suppressing the production of the emission.
\end{abstract}




\section{Introduction}\label{sec:intro}
Hot magnetic stars are early-type stars (spectral types OBA) with large-scale surface magnetic fields of $\sim$kG strength that are extremely stable \citep[e.g.][]{shultz2018}. Their radiatively driven winds, and large-scale \citep[usually dipolar, e.g.][]{kochukhov2019} magnetic fields interact with each other, leading to the formation of co-rotating magnetospheres \citep[e.g.][]{petit2013}. Inside the magnetosphere, a rich variety of electromagnetic phenomena take place that give rise to emission from X-ray to radio bands
\citep[e.g.][]{drake1987,trigilio2000,oksala2012,petit2013,eikenberry2014,naze2014}.
The high stability of the magnetic fields and their relatively simple topologies, as well as the precisely measured magneto-rotational properties \citep[e.g.][]{shultz2018} make these ideal laboratories to investigate the various electromagnetic phenomena resulting from the wind-field interplay. This might have implications for objects of much later spectral types, but also characterized by the presence of large-scale surface magnetic fields such as the ultracool dwarfs or UCDs, \citep[e.g.][and references therein]{kochukhov2021}, and possibly, also in brown dwarfs \citep[e.g.][]{berdyugina2017}. Evidence for this possibility was provided by \citet{leto2021}, who showed that the empirical relation between the incoherent gyrosynchrotron radio luminosity and stellar parameters obtained for hot magnetic stars is valid even for ultracool dwarfs and Jupiter. 


One of the attributes observed from large-scale magnetospheres is the generation of coherent radio emission by the process of electron cyclotron maser emission (ECME). ECME is a highly directed emission with intrinsically narrow bandwidth. The frequency of emission is proportional to the local electron gyrofrequency $\nu_\mathrm{B}\propto B$ ($B$ is the local magnetic field strength). As a result, the higher frequencies are produced closer to the star (where $B$ is stronger). For the hot magnetic stars, the direction of emission is nearly perpendicular to the magnetic dipole axis and the local magnetic field direction \citep[the tangent plane beaming model,][]{trigilio2011}, which results in the fact that they are observed 
near the rotational phases where the stellar disk-averaged line-of-sight magnetic field \bz~is zero (hereafter referred to as magnetic nulls). This implies that even if a star produces ECME, the radiation may not be observable if the star's magnetic nulls are not observable, which is determined by the orientation of the magnetic axis with respect to the line of sight and the rotation axis. Assuming an axi-symmetric dipolar magnetic field, this translates to the geometrical condition for detectability: $i+\beta\geq 90^\circ$, where $i$, called the inclination angle, is the angle between the star's rotation axis and the line-of-sight; and $\beta$, called the magnetic obliquity (which will henceforth be referred as simply `obliquity'), is the angle between the star's rotational axis and magnetic dipole axis.  

The first hot magnetic star discovered to produce ECME was CU\,Vir \citep{trigilio2000}. Highly circularly polarized, periodic radio pulses (much brighter than the incoherent radio emission) were observed at 1.4 GHz near the magnetic nulls, but were absent at and above 5 GHz \citep{trigilio2000}. The high brightness temperature, high circular polarization, and coincidence with the magnetic nulls confirmed that the emission mechanism is ECME. Recently, however, it has been demonstrated by \citet{das2020a}, \citet{das2021} and \citet{das2022} that high observed circular polarization is not a necessary criterion for the pulses to be of ECME origin. 
Further, the study of the ECME from CU\,Vir at frequencies below 1 GHz \citep{das2021} gave evidence that the maser emission can occur at rotational phases far from the magnetic nulls. Such evidence suggested that on the low-frequency side of the ECME spectrum the maser radiation emission beam is wider than what is expected following the tangent plane beaming model \citep{trigilio2011}. Therefore, the ECME detection probability becomes consequently higher when early-type magnetic stars are observed at low radio frequencies.

Following the discovery of coherent radio emission from CU\,Vir, no other similar star was reported for more than a decade, until \citet{chandra2015} reported the second candidate (HD\,133880), which was later confirmed by \citet{das2018}. The discovery of just two stars within a span of more than a decade led to the impression that the generation of ECME is rare amongst hot magnetic stars. This was unsatisfactory since they seemingly have all the necessary `ingredients' for the production of ECME. Between 2018 and 2021, however, five more such stars were reported \citep{lenc2018,leto2019,das2019b,das2019a,leto2020,leto2020b}. In 2022, \citeauthor{das2022} reported eight more such stars. Based on this large number of discoveries, \citet{das2022} estimated that more than 32\% of hot magnetic stars are capable of producing ECME. This fraction is extremely conservative, since the phenomenon has not been searched for in the majority of the population. The large value of the conservative lower limit to the incidence fraction of ECME rules out the notion that the phenomenon is rare amongst hot magnetic stars \citep{das2022}. The hot magnetic stars capable of producing ECME have recently been given the designation of `Main-sequence Radio Pulse emitters' \citep[MRPs,][]{das2021}.


The idea that ECME is probably ubiquitous among magnetic early-type stars bears significant importance due to the fact that this emission contains a treasure-trove of information about the stellar magnetosphere. 
The importance of ECME is well-recognized in the case of low-mass stars as well as brown dwarfs. Such emission was used to detect kG strength magnetic field in ultracool dwarfs and brown dwarfs \citep[e.g.][]{hallinan2007, kao2018}, to infer magnetic field configurations \citep[e.g.][]{llama2018,davis2021}, and also to detect the presence of planetary companions \citep[e.g.][]{kavanagh2022,perez-torres2021}. 
For the magnetic massive stars, the importance of ECME lies in the fact that it is the only kind of magnetospheric emission that is sensitive to small-scale changes in the stellar magnetosphere (due to its highly directed nature). 
\citet{das2020b} showed via simulation that ECME pulse characteristics are strongly affected by the magnetospheric plasma density distribution, and \citet{das2021} suggested that ECME can give us information about small-scale magnetospheric explosions, which are responsible for regulating plasma flow in stellar magnetospheres \citep[centrifugal breakout,][]{shultz2022,owocki2022}. Note that other magnetospheric emissions such as H$\mathrm{\alpha}$ emission, are also sensitive to the overall plasma distribution in the magnetosphere, and the variation of their emission profiles can be well-explained in the framework of the rigidly rotating magnetosphere model \citep[RRM model,][]{townsend2005,ud-doula2008,shultz2020,owocki2020}. However, none of these diagnostics are directed emission, and hence lack the ability to provide `high spatial-resolution' information. From this perspective, it is important to characterize the ECME phenomenon so that it can be used as a probe complementary to existing diagnostics (H$\mathrm{\alpha}$, X-ray emission etc.). To achieve that, one of the essential steps is to identify the stellar physical parameters that drive the phenomenon. This was first attempted by \citet{das2022}.
By using fourteen of the fifteen known hot magnetic stars producing ECME
\footnote{Although there were 15 hot magnetic stars known to produce ECME, one of them \citep[$\rho\,\mathrm{Oph\,C}$,][]{leto2020b} is such that its ECME is always visible so that the contribution of incoherent emission to the observed flux density could not be estimated. The data for this star was hence not used in the analysis of \citet{das2022}.}, 
they derived the first empirical relation connecting the maximum observed ECME luminosity and stellar parameters. The parameters that they found to exhibit the strongest correlation with ECME luminosity are the maximum surface magnetic field strength ($B^0_\mathrm{max}$) and the stellar effective temperature ($T_\mathrm{eff}$). They constructed a quantity called `$X$' (we will call it the `$X$-factor'), given by the following equation:
\begin{align}
    X=\frac{B^0_\mathrm{max}}{{(T_\mathrm{eff}-16.5)}^2} \label{eq:x_factor}
\end{align}
where $B^0_\mathrm{max}$ is in kG and $T_\mathrm{eff}$ is in kK, with which the ECME luminosity was found to vary almost linearly. 
Thus, they surprisingly found that the phenomenon is favoured for stars with $T_\mathrm{eff}$ around 16--17 kK, a value for which there is not at present a theoretical justification. 
\citet{das2022} obtained a nearly linear equation between the peak ECME luminosity \Le~(calculated from $S_\mathrm{peak}d^2$, where $S_\mathrm{peak}$ is the observed peak flux density at either circular polarization in excess of the basal gyrosynchrotron emission, and $d$ is the distance to the star) and the $X$-factor: \Le$\propto X^\mathrm{0.8\pm0.1}$.

A correlation of radio luminosity with the magnetic field strength has also been observed for incoherent gyrosynchrotron emission \citep{leto2021,shultz2022}, however, no correlation with temperature was found for the incoherent radio emission. This was noted by \citet{shultz2022} who pointed out that for the incoherent gyrosynchrotron emission, a strong correlation between the 5 GHz radio luminosity and the temperature was also initially proposed \citep{linsky1992}, but this now seems to have been entirely an artifact of small-sample analysis. Thus, for the coherent radio emission, before attempting to interpret the empirical relation, it is important to test its robustness against the sample size. Besides, the incoherent radio luminosity was also found to correlate with the stellar rotation period \citep{leto2021,shultz2022}, however, no such correlation was observed for ECME. \citet{das2022}, however, noted that the MRP sample essentially spans only a very narrow range of stellar rotation periods (0.5--2 days, with just one star with rotation period $\approx 4$ days), whereas the longest rotation period among the `radio-bright' (in terms of incoherent radio emission) hot magnetic stars has been found to be $\approx 10$ days \citep{landstreet2000,shultz2022}. 
This makes it practically impossible to investigate if there is any dependence of ECME luminosity on rotation period. 

In an attempt to test the predictive capability of the empirical relation suggested by \citet{das2022}, we observed five hot magnetic stars at sub-GHz frequencies.
Four of these stars are predicted to produce detectable ECME, and one is predicted to be incapable of producing it. In this paper, we report the results of these observations, and their implication for the relation among ECME luminosity, stellar parameters, and the incoherent radio luminosity.

This paper is structured as follows: in the next section (\S\ref{sec:targets}), we describe our sample of stars. This is followed by a description of the radio observations and data reduction (\S\ref{sec:data}). We present the resulting light curves in \S\ref{sec:results}, and discuss some key findings in \S\ref{sec:discussion}. We finally summarize our conclusions in \S\ref{sec:summary}.


\begin{table*}
\begin{threeparttable}
{\tiny
\caption{The spectral type dipole strength $B^0_\mathrm{max}$, effective temperature $T_\mathrm{eff}$, stellar radius $R_*$, inclination angle $i$, obliquity $\beta$ and distance $d$ for our sample of stars. Also shown are the reference heliocentric Julian day $\mathrm{HJD_0}$ and rotation period $P_\mathrm{rot}$ used to phase the data. Note that HD\,37479 is known to exhibit detectable rotation period evolution, and we have used a variable period ephemeris for this star (\S\ref{sec:targets}), the listed $P_\mathrm{rot}$ for the star corresponds to the instantaneous rotation period at the given epoch. In addition to the stellar parameters, we have also listed the approximate values of the $X$-factor (Eq. \ref{eq:x_factor}), and the corresponding predicted \textit{excess} peak flux density at $\approx 0.6$ GHz, $S_\mathrm{peak}$. The last column ($S_\mathrm{basal}$) shows the existing measurements for the incoherent flux densities of these stars over 0.6--0.8 GHz. Note that the averaging times for the different $S_\mathrm{basal}$ varies from $\approx 1$ hours to $\approx 4.3$ hours (equivalently, 0.05 rotation cycles to 0.14 rotation cycles).
\label{tab:star_properties}}
\begin{tabular}{rcccllllllccc}
\hline\hline
HD &Sp. &$B^0_\mathrm{max}$ & $T_\mathrm{eff}$& $R_*$ & $i$& $\beta$ & $d$ &$\mathrm{HJD_0}$ & $P_\mathrm{rot}$ & $X$ & ${S_\mathrm{peak}}^*$ & ${S_\mathrm{basal}}^{**}$ \\
No.& type & (kG) & (kK) & ($R_\odot$) & (degree)& (degree) & (pc) & & (days) & & (mJy) & (mJy) \\
\hline
35502 & B5IV$^{14}$ & ${7.3\pm 0.5}^{[1]}$ & ${18.4\pm 0.6}^{[2]}$ & ${2.96\pm 0.10}^{[1]}$ &${25.6\pm1.1}^{[1]}$ & ${70\pm1}^{[1]}$ & ${383\pm 9}^{[7]}$ & 2456295.812850 & ${0.853807(3)}^{[2]}$ & $2.02$ & $\sim 2.5$ & ${0.64\pm 0.16}^{[12]}$ \\
36526 & B8II$^{15}$ & ${11.2\pm 0.5}^{[1]}$ & ${15\pm 2}^{[3]}$ & ${2.35^{+0.09}_{-0.12}}^{[1]}$& ${46\pm2}^{[1]}$& ${57\pm 2}^{[1]}$ & ${410\pm 14}^{[7]}$ & 2455611.93 & ${1.54185(4)}^{[8]}$ & 4.98 & $\sim 4.6$ & ${<0.1}^{[12]}$ \\
37479 & $^{14}$ & ${10^{+2}_{-1}}^{[1]}$ & ${23\pm 2}^{[3,4]}$ & ${3.39^{+0.04}_{-0.06}}^{[1]}$ & ${77\pm4}^{[1]}$ & ${38\pm 9}^{[1]}$ & ${438\pm 18}^{[7]}$ & 2459580.44769 & ${1.1908675(30)}^{[9]}$  & 0.24 & $\sim 0.3$ & ${1.25\pm 0.28}^{[13]}$ \\
61556 & $^{16}$ & ${2.8\pm 0.3}^{[1]}$ & ${18.5\pm 0.8}^{[5]}$ & ${3.3\pm 0.6}^{[1]}$ & ${35\pm 7}^{[1]}$ & ${59\pm 6}^{[1]}$ & ${114\pm 3}^{[7]}$ & 2455199.1 & ${1.90871(7)}^{[5]}$ & 0.7 & $\sim 10$ & ${1.21\pm 0.18}^{[12]}$ \\
182180 & B2V$^{16}$ & ${9.5\pm 0.6}^{[1]}$ & ${19.8\pm 1.4}^{[6]}$& ${3.2\pm0.1}^{[1]}$ &${53^{+9}_{-5}}^{[1]}$ & ${82\pm 4}^{[1]}$& ${233\pm 6}^{[7]}$ & 2454940.830 & ${0.5214404(6)}^{[10,11]}$ & 0.87 & $\sim 3.4$ & ${4.17\pm 0.42}^{[12]}$ \\
\hline
\end{tabular}
\begin{tablenotes}
       \item $^*$ Predicted
      \item $^{**}$ Observed
      \item Reference key: 1, \citet{shultz2019c}; 2, \citet{sikora2016}; 3, \citet{shultz2019b}; 4, \citet{hunger1989}; 5, \citet{shultz2015}; 6, \citet{rivinius2013}; 7, \citet{gaia2018}; 8, \citet{shultz2018}; 9, Petit et al. (in prep.); 10, \citet{oksala2010}; 11, \citet{rivinius2010}; 12, \citet{shultz2022}; 13, \citet{chandra2015}; 14, \citet{abt1977}; 15, \citet{houk1999}; 16, \citet{houk1982}.
\end{tablenotes}
}
\end{threeparttable}
\end{table*}

\section{Targets}\label{sec:targets}
The five stars that we chose to observe are HD\,35502, HD\,36526, HD\,37479, HD\,61556 and HD\,182180. The physical properties of the five stars are given in Table \ref{tab:star_properties}. All five of them are classified as having centrifugal magnetospheres (CMs) by \citet{petit2013}, which means that the extent of their magnetospheres are larger than the distance at which the centrifugal force due to co-rotation balances gravity. 

HD\,35502 is a hierarchical triple system with the primary B star as the magnetic component; the secondary components are two A-type stars that orbit the primary with an orbital period of over 40 years \citep{sikora2016}. In this work, by HD\,35502, we will refer to the system's magnetic component.

Among our sample of stars, magnetospheric emission in the optical and X-ray has been detected from HD\,35502 \citep{sikora2016, petit2013}, HD\,37479 \citep[e.g.][]{oksala2012,petit2013} and HD\,182180 \citep{oksala2010,rivinius2010,rivinius2013,naze2014}. In fact HD\,37479 (commonly known as $\sigma\,\mathrm{Ori\,E}$) was the first star in which the H$\mathrm{\alpha}$ emission was inferred to be a magnetospheric signature \citep{landstreet1978}. Among the remaining two stars, magnetospheric emission were searched in both HD\,61556 \citep[in optical and UV lines,][]{shultz2015}, and HD\,36526 \citep[in optical,][]{shultz2020}, but no detection was obtained in either case.

Incoherent radio emission has been previously detected from HD\,35502 \citep[from 0.6 up to 15 GHz,][]{drake2006, leto2021, shultz2022}, HD\,37479 \citep[from 0.6 up to 43 GHz, ][]{drake1987,linsky1992,leone1996,leone2004,leto2012,chandra2015}, HD\,61556 \citep[at 0.6 and 1.5 GHz,][respectively]{pritchard2021,shultz2022} and HD\,182180 \citep[from 0.6 up to 43 GHz,][]{leto2017,shultz2022}. HD\,36526, on the other hand, has never been detected in radio. The $3\sigma$ upper limit to its flux density at 0.6 GHz is 0.10 mJy \citep{shultz2022}, whereas at 5 GHz, the upper limit is 0.18 mJy \citep{linsky1992}. 

In Table \ref{tab:star_properties}, we also list the value of the $X$-factor (Eq. \ref{eq:x_factor}). The peak ECME flux densities (at around 0.6 GHz) predicted from the empirical relation are listed in Table \ref{tab:star_properties}. Note that the values listed correspond to the excess flux density due to ECME, and not the total flux density. We find that apart from HD\,37479, the predicted ECME enhancement in flux density is much larger than the error bars associated with the basal flux densities, so that if they indeed produce ECME, the signature will be easily detectable. HD\,37479, on the other hand, has a predicted enhancement that is comparable to the error bar in its basal flux density making it unlikely that the signature of ECME (even if it produces the emission) will be detectable. Thus, to summarize, the empirical relation proposed by \citet{das2022} predicts that ECME will be detected from all targets but HD\,37479. 
{\color{black}
Excluding HD\,37479, which is known to be undergoing a constant spin-down \citep[e.g.][]{townsend2010,mikulasek2016}, the data were phased using the following equation:
\begin{align}
    \vartheta&=\frac{\mathrm{HJD}-\mathrm{HJD_0}}{P_\mathrm{rot}}, \label{eq:linear_ephem}
\end{align}
where the rotational phase is given by the fractional part of $\vartheta$. $\mathrm{HJD_0}$ and $P_\mathrm{rot}$ are listed in Table \ref{tab:star_properties}.

In case of HD\,37479, we adopted a variable period ephemeris presented by Petit et al. (in prep.), which accounts for new information, including TESS photometry, and which incorporates a double derivative term in contrast to the previous, steady spin-down ephemeris of \citet{townsend2010}:
}
\begin{align}
    \vartheta(t)&=\vartheta_0-\frac{\dot{P}_\mathrm{rot}\vartheta_0^2}{2}-\frac{P_\mathrm{rot}\ddot{P}_\mathrm{rot} {\vartheta_0}^3}{6},\,\,\,\,\mathrm{where\,}\vartheta_0=\frac{\mathrm{HJD}-\mathrm{HJD_0}}{P_\mathrm{rot}}\label{eq:non_linear_ephem}
\end{align}
$P_\mathrm{rot}=1.1908677(3)$ days is the instantaneous period at the epoch $\mathrm{HJD}_0=2459580.4478(3)$, $\dot{P}_\mathrm{rot}=2.29(15)\times 10^{-9}$
and $\ddot{P}_\mathrm{rot}=-87(20)\times 10^{-15}\,\mathrm{day^{-1}}$. The rotational phase is then given by the fractional part of $\vartheta(t)$.

\section{Observation and data analysis}\label{sec:data}
Our aim was to verify whether HD\,35502, HD\,36526, HD\,61556 and HD\,182180 produce detectable ECME, and to confirm that HD\,37479 does not produce detectable ECME. We hence adopt two different observing strategies for the first four targets, and for HD\,37479. The former were observed around/near one of their magnetic nulls. HD\,61556 was observed as part of an earlier program in the year 2018, and it was observed over a rotational phase window of width 0.1 cycles only \citep[in addition to another observation of similar duration around a phase much offset from a magnetic null, reported in ][included in this paper]{shultz2022}. On the other hand, the other three stars were observed over a rotational phase window with a width of 0.4 cycles. In the case of HD\,37479, since we want to check if it is a non-MRP, we observed for one full rotation cycle, as otherwise, a non-detection of ECME over a limited rotational phase range always leaves the possibility of the presence of the emission at a rotational phase outside the covered range.

We observed our targets (other than HD\,61556) with the upgraded Giant Metrewave Radio Telescope \citep[uGMRT,][]{gupta2017,swarup1991} using its band 4 (550--900 MHz) from November 2021 through March 2022. In addition, we also analyzed archival GMRT data for HD\,37479 and HD\,182180. The details of these data are given in Table \ref{tab:obs}.

\begin{table}
\begin{threeparttable}
{\scriptsize
\caption{Log of GMRT observations of our targets. All observations except for the one acquired in the year 2015 (last row) were acquired using the uGMRT, whereas the one in the year 2015 was acquired with the legacy GMRT.\label{tab:obs}}
\begin{tabular}{ccccll}
\hline\hline
HD & Date & HJD range & Eff. band & Flux & Phase\\
No. & of Obs. & $-2.45\times10^6$ & (MHz) & Calibrator  & Calibrator\\
\hline\hline
35502 & 2022-01-16 & $9596.20\pm 0.19$ & $570-804$ & 3C48,& J0503+020,\\
& & & & 3C147 & J0607--085\\
\hline
36526 & 2021-12-31 & $9580.25\pm 0.19$ & $570-804$ & 3C48 & J0503+020,\\
& & & & & J0607--085\\
& 2022-02-15 & $9626.12\pm 0.18$ & $570-804$ & 3C48, & J0503+020,\\
& & & & 3C147 & J0607--085\\ 
\hline
37479 & 2021-11-08 & $9527.39\pm 0.18$ & $570-804$ & 3C48,& J0503+020,\\
& & & & 3C286 & J0607--085\\
& 2021-11-16 & $9535.41\pm 0.14$ & $570-804$ & 3C48, & J0607--085\\
& & & & 3C286 & \\
& 2021-11-30 & $9549.39\pm 0.12$ & $570-804$ & 3C48, & J0607--085\\
& & & & 3C286 & \\ 
& 2022-01-07 & $9587.20\pm 0.15$ & $570-804$ & 3C48,& J0503+020,\\
& & & & 3C286 & J0607--085\\
& 2018-05-18 & $8256.96\pm 0.03$& $570-667$ & 3C286 & J0607--085\\
& 2018-06-09 & $8278.86\pm 0.02$  & $560-725$ & 3C48 & J0607--085\\
\hline
61556 & 2018-05-04 & $8243.00\pm 0.05$ & $558-680$ & 3C48, & J0837-198\\
& & & & 3C286 & \\
& 2018-06-04 \tnote{1}& $8273.92\pm 0.06$& $570-726$ & 3C48 & J0735-175\\
\hline 
182180 & 2022-03-04 & $9642.66\pm 0.11$ & $570-804$ & 3C286 & J1924--292\\
& 2018-04-30 & $8238.54\pm 0.01$ & $560-677$& 3C48 & J1924-292\\
& 2018-08-17 & $8348.30\pm 0.02$ & $560-726$& 3C48 & J1924-292\\
& 2015-09-24 & $7290.17\pm 0.05$ & $595-619$ & 3C48 & J1830-360\\
\hline 
\end{tabular}
}
\begin{tablenotes}
      \item[1] The average flux density (total intensity) was reported in \citet{shultz2022}.
\end{tablenotes}
\end{threeparttable}
\end{table}

The data were analyzed using the Common Astronomy Software Applications \citep[\textsc{casa},][]{mcmullin2007} following the procedure described by \citet{das2019b,das2019a}.

Throughout this paper, we use the IAU/IEEE convention to define right and left circular polarization (RCP and LCP respectively), and to define Stokes V.

\section{Results}\label{sec:results}
Following \citet{das2022}, the only condition that we impose to identify an MRP candidate is the observation of a significant flux density enhancement close to a magnetic null. \citet{das2022} provided a minimum flux density gradient condition to quantify the criterion of significant flux density enhancement, which is that the rotational phase range ($\Delta \phi_\mathrm{rot}$) over which the enhancement reaches its peak value from its basal value, should be smaller than $\Delta\phi_\mathrm{crit}=$0.159 \citep[see Eq. 2 of][]{das2022}. Note that this criterion assumes that the star's \bz~variation with rotational phase can be approximated as a sinusoid, and that the incoherent radio flux density follows the rotational modulation of \bz~and hence varies as 
$\sin^2(2\pi\phi_\mathrm{rot})$, where $\phi_\mathrm{rot}$ represents the rotational phase. This is however an extremely idealized situation, since very often, \bz~does not vary sinusoidally, and more complex modulations are seen in both \bz~and that of the incoherent radio flux density \citep[e.g. HD\,133880,][]{lim1996,bailey2012,kochukhov2017}. This makes the exact value of $\Delta\phi_\mathrm{crit}$ much less useful while investigating whether or not an enhancement can be attributed to ECME. But the basic idea,
which is that the enhancement due to ECME should occur over a much smaller time-scale (rotational phase range) than that expected for incoherent emission, remains valid.
In order to examine whether or not an enhancement is due to ECME, we consider $\Delta\phi_\mathrm{crit}$, as well as other indicators, such as brightness temperature, magnitude of the enhancement, and the percentage of circular polarization.

We present the results obtained for each star in the following subsections. Along with their radio light curves, we also show their \bz~variation with rotational phase $\phi_\mathrm{rot}$. We fit the \bz~curve using the following equation:
\begin{align}
    \langle B_\mathrm{z}\rangle&=\sum_{n=0}^{N} B_n\sin2\pi (n\phi_\mathrm{rot}+\phi_n)\label{eq:bz_curve}
\end{align}

The integer $N$ is chosen based on the complexity of the observed variation.

\subsection{HD\,35502}\label{subsec:hd35502}
\begin{figure}
    \centering
    \includegraphics[width=0.45\textwidth]{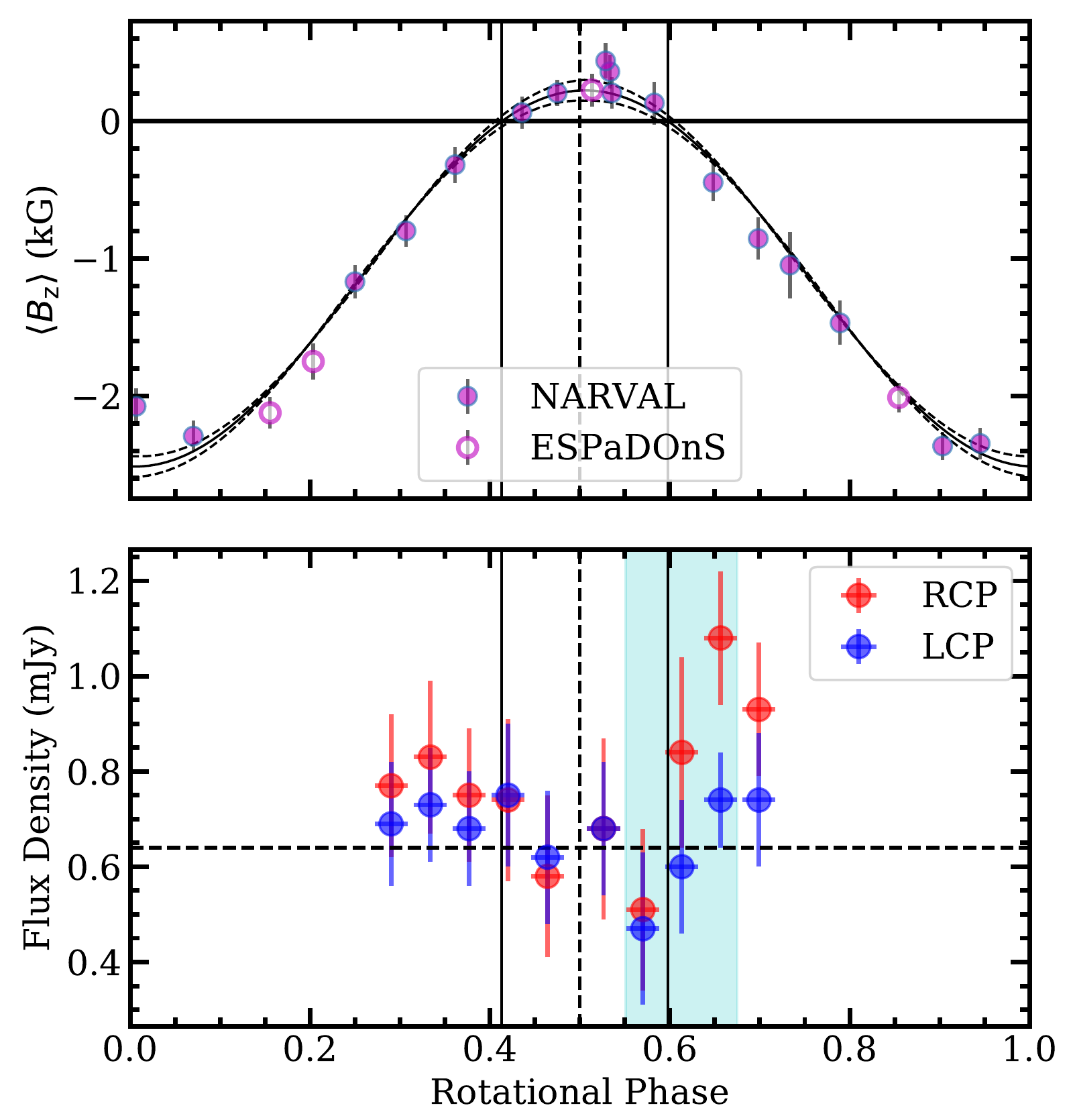}
    \caption{\textit{Top:} The \bz~curve of HD\,35502. The measurements were originally reported by \citet{sikora2016}, which were later revised (binary disentanglement) by \citet{shultz2018}. The magnetic null phase was at around 0.5 (vertical dashed line) according to \citet{sikora2016}, but the revised study found that there are two magnetic nulls at $0.41\pm0.02$ and $0.60\pm0.02$ phases (solid vertical lines). The fitted curve corresponds to Eq. \ref{eq:bz_curve} with $n=1$. The radio measurements were acquired around 0.5 phase. \textit{Bottom:} The light curves of HD\,35502 at 687 MHz. The integration time corresponding to each data point is 45 minutes or 0.04 rotation cycle of the star. The dashed horizontal line corresponds to the basal flux density listed in Table \ref{tab:star_properties}.}
    \label{fig:hd35502}
\end{figure}

HD\,35502 is one of the stars that is predicted to produce detectable ECME by the empirical relation of \citet{das2022}. The $X$-factor for this star lies between 1.1 and 4.6 after we take into account the error bars in the measurements of its dipole strength and surface temperature. The corresponding peak flux density lies between $\sim 1$ and $8$ mJy (after also considering the uncertainty in the empirical relation). The longitudinal magnetic field of the star is predominantly negative (top panel of Figure \ref{fig:hd35502}). Initially, the star was thought to have just one single magnetic null at phase $\approx 0.5$ \citep{sikora2016}.
However, upon 
disentangling the line profiles of the multiple system's 3 components, the star's  \bz~curve was found to exhibit two closely spaced magnetic nulls at phases $\approx 0.4\,\mathrm{and}\, 0.6$ \citep[top panel of Figure \ref{fig:hd35502},][]{shultz2018}.

Despite the refinement of the star's magnetic null phases, we scheduled our radio observation around phase 0.5. Note that for the star HD\,37017, there also exist two closely spaced magnetic nulls in its \bz~curve, but ECME was observed at a phase that lies between the two magnetic nulls \citep{das2022}.

The bottom panel of Figure \ref{fig:hd35502} shows the radio light curves at 687 MHz.
Contrary to the prediction, we do not detect any clear signature of ECME. 
{\color{black} The maximum observed flux density is $1.1\pm0.1$ mJy.}
There is some indication of enhancement in RCP flux density between phases 0.55--0.70 (shaded region in Figure \ref{fig:hd35502}). The corresponding value of $\Delta \phi_\mathrm{rot}$ is $0.124$. Although this satisfies the necessary criterion to attribute an enhancement to ECME, the error bars associated with the relevant flux density measurements are too large to confirm it. Within the rotational phase range covered by our observation, the light curve exhibits a variability with a timescale of $\Delta\phi_\mathrm{rot}\sim 0.2$, which we attribute to rotational modulation of gyrosynchrotron emission

In \S\ref{subsec:no_ecme_hd35502}, we discuss why this star could be a non-MRP despite the prediction of the empirical relation of \citet{das2022}.

\subsection{HD\,36526}\label{subsec:hd36526}
\begin{figure}
    \centering
    \includegraphics[width=0.45\textwidth]{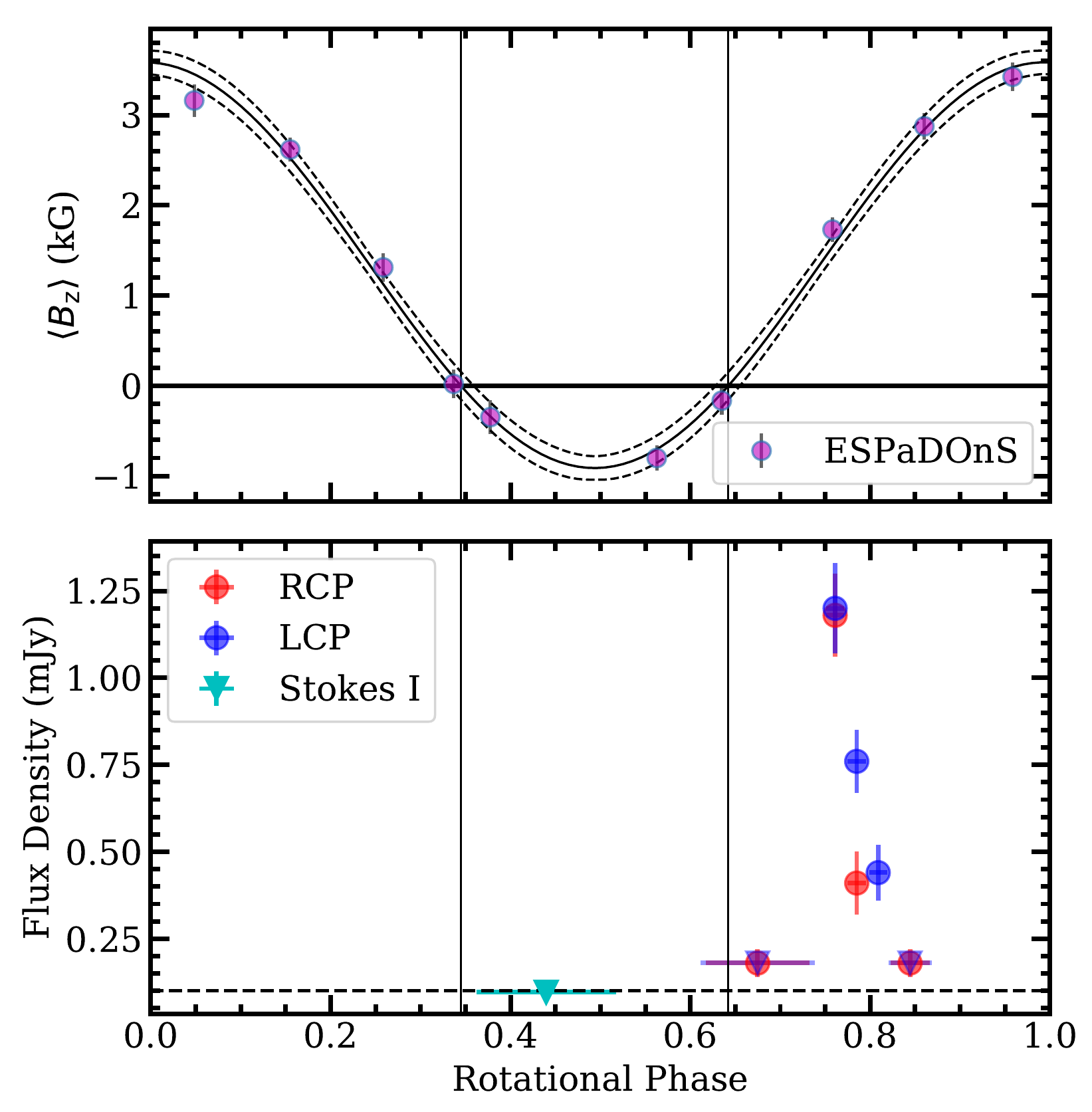}
    \caption{\textit{Top:} The variation of \bz~with rotational phase of HD\,36526 \citep{shultz2018}. By fitting a sinusoid ($n=1$ in Eq. \ref{eq:bz_curve}), the magnetic nulls were found to lie at phase $0.34\pm0.01$ and $0.64\pm0.01$ (marked with black vertical lines). \textit{Bottom:}
     The light curves at 687 MHz around its magnetic null at phase $\approx 0.64$. The smallest averaging time is 45 minutes, or 0.02 rotation cycles. The triangles represent the $3\sigma$ upper limit to the flux density. The dashed horizontal line corresponds to the $3\sigma$ upper limit to the basal flux density listed in Table \ref{tab:star_properties}.}
    \label{fig:hd36526}
\end{figure}
HD\,36526 is another star for which detectable ECME (at our observing frequency) was predicted by the empirical relation of \citet{das2022}. Due to the relatively large uncertainty associated with its effective temperature ($T_\mathrm{eff}=15\pm2$ kK, Table \ref{tab:star_properties}), the corresponding uncertainty in its $X$-factor is also large, especially because the latter includes the term ${(T_\mathrm{eff}-16.5)}^2$ in its denominator. The range of possible values of is $X$-factor is 0.9--47, and the corresponding peak flux density (after we also take into account the uncertainty associated with the empirical relation itself) lies between $\sim 1$ and $50$ mJy. Since the lower limit to the predicted peak flux density is already much higher than the existing estimate for its basal flux density at 687 MHz (the 3$\sigma$ upper limit is $0.1$ mJy, Table \ref{tab:star_properties}), the star is predicted to be an MRP by the empirical relation of \citet{das2022}.

Figure \ref{fig:hd36526} shows the light curves of this star acquired around its magnetic null at rotational phase $\approx 0.64$. Although we had obtained a near continuous rotational phase coverage of the star over phase $\approx 0.40-0.85$, a large amount of data were found to be corrupted by Radio Frequency Interference (RFI) leading to the gap in the light curve between phases 0.5 and 0.6. We did not detect the star over rotational phases 0.36--0.52, and the $3\sigma$ upper limit to the total intensity comes out to be $0.096$ mJy, consistent with the existing estimate of the star's basal flux density \citep{shultz2022}. Between phases 0.62 and 0.86, the star was detected at high significance, {\color{black} with the maximum flux density reaching $1.2\pm 0.1$ mJy at phase 0.76}. The corresponding lower limit to the brightness temperature, obtained by setting the emission site equal to the stellar disk, is $>10^{12}$ K, confirming the coherent nature of the emission. The high brightness temperature, order of magnitude enhancement in flux density, and confinement to a narrow rotational phase window ($\Delta\phi_\mathrm{rot}\sim 0.1$) near a magnetic null, confirm that the enhanced emission is due to ECME, despite the fact that the net circular polarization is nearly zero \citep[also observed for several other MRPs, e.g. HD\,12447,][]{das2022}.

The excess peak flux density $S_\mathrm{peak}$ is $1.1<S_\mathrm{peak}<1.2$ mJy, consistent with the prediction of the empirical relation of \citet{das2022}. Note that this is the first detection of magnetospheric emission from this star.

\subsection{HD\,37479}\label{subsec:hd37479}
\begin{figure}
    \centering
    \includegraphics[width=0.45\textwidth]{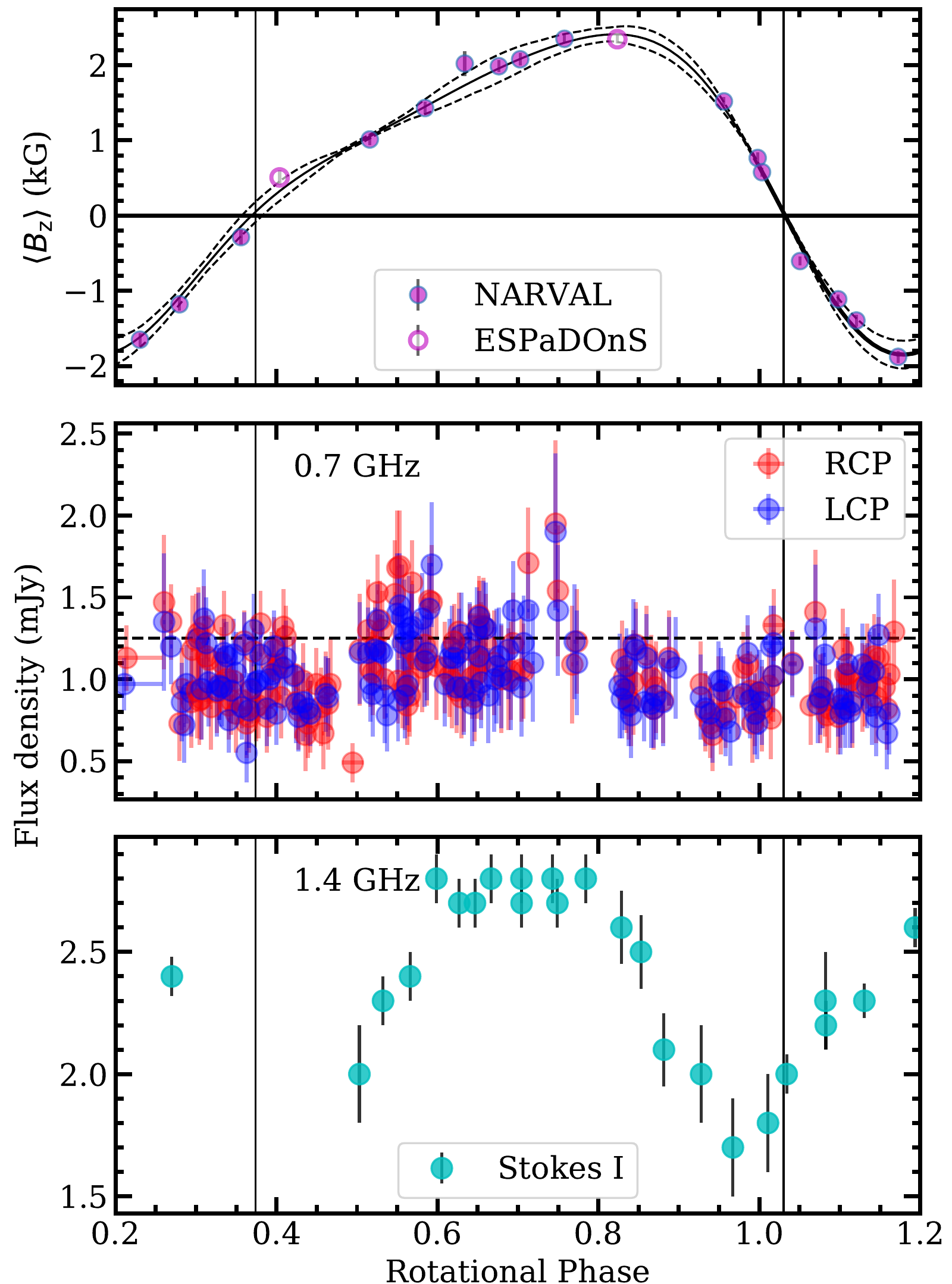}
    \caption{\textit{Top:} The variation of \bz~with the rotational phase of HD\,37479. The measurements were reported by \citet{oksala2012}. The solid line fitted to the data corresponds to Eq.\ref{eq:bz_curve} with $n=3$. The magnetic nulls are at phases $0.03\pm0.01$ and $0.36\pm0.02$ (marked with the black vertical lines).
    \textit{Middle:}
    The light curves of HD\,37479 at 687 MHz. The smallest averaging time is 45 minutes, or 0.02 rotation cycle. The dashed horizontal line corresponds to the basal flux density listed in Table \ref{tab:star_properties}.
    \textit{Bottom:}
    The total intensity light curves of the star at 1.4 GHz as reported by \citet{leto2012}.
    }
    \label{fig:hd37479}
\end{figure}
HD\,37479 (more commonly known as $\sigma\,\mathrm{Ori\,E}$) is the only star in our sample that is predicted to be a non-MRP by \citet{das2022}. 
Note that in this context, the term `non-MRP' signifies the lack of an observable signature of the ECME phenomenon (since the predicted ECME flux density is smaller than the incoherent flux density, see Table \ref{tab:star_properties}), which does not necessarily imply the absence of the phenomenon in the stellar magnetosphere.
The star was observed for nearly a complete rotation cycle over 1.4--15 GHz by \citet{leto2012} specifically to search for ECME (with their lowest frequency lightcurve reproduced in the bottom panel of Figure \ref{fig:hd37479}). No signature of ECME was detected, which was attributed to the presence of high-order component(s) in the stellar magnetic field (manifested as non-sinusoidal variation of \bz~with stellar rotational phase, top panel of Figure \ref{fig:hd37479}). However, as shown by \citet{das2022}, a number of MRPs possess magnetic fields more complex than a simple dipole, and yet produce detectable ECME, making the role of the complex magnetic field less obvious in suppressing ECME.

The $X$-factor of HD\,37479 is only 0.24, due to its relatively high surface temperature. As mentioned already in \S\ref{sec:targets}, the predicted excess peak flux density is comparable to the error bars in the basal flux density, which means that the star should not be an MRP (according to the empirical relation).
To test this prediction, we acquired the near-full rotation cycle observation  of the star at 687 MHz. The result is shown in Figure \ref{fig:hd37479}. There is no signature of ECME (no strong enhancement in flux density), consistent with the prediction. 
The maximum observed flux density of $2.0\pm0.5$ mJy is consistent with the basal flux density within error bars. 
This makes HD\,37479 the only hot magnetic star confirmed to be a non-MRP down to sub-GHz frequencies.

Figure \ref{fig:hd37479} also shows the sharp decrease in the amplitude of rotational modulation of the incoherent radio emission as the frequency of emission decreases from 1.4 GHz to 0.7 GHz. Such an effect is expected as the area of the emission site becomes less and less concentrated with the decrease in observation frequency \citep[e.g. see Figure 4 of][]{leto2021}.
The incoherent flux density also decreases significantly from 1.4 GHz to 0.7 GHz. Based on the radio spectrum \citep[Figure 2 of ][]{leto2021}, the extrapolated incoherent flux density of HD\,37479 at 300 MHz is $\sim 0.1$ mJy. If the star indeed produces ECME and the peak flux density remains roughly constant with further decrease in frequency, it might be possible to detect its signature by observing the star at frequencies of $\sim 300$ MHz or below. Thus lower frequency follow-up observations will be able to provide further insight towards the `non-MRP behavior' of the star.

\subsection{HD\,61556}\label{subsec:hd61556}
\begin{figure}
    \centering
    \includegraphics[width=0.45\textwidth]{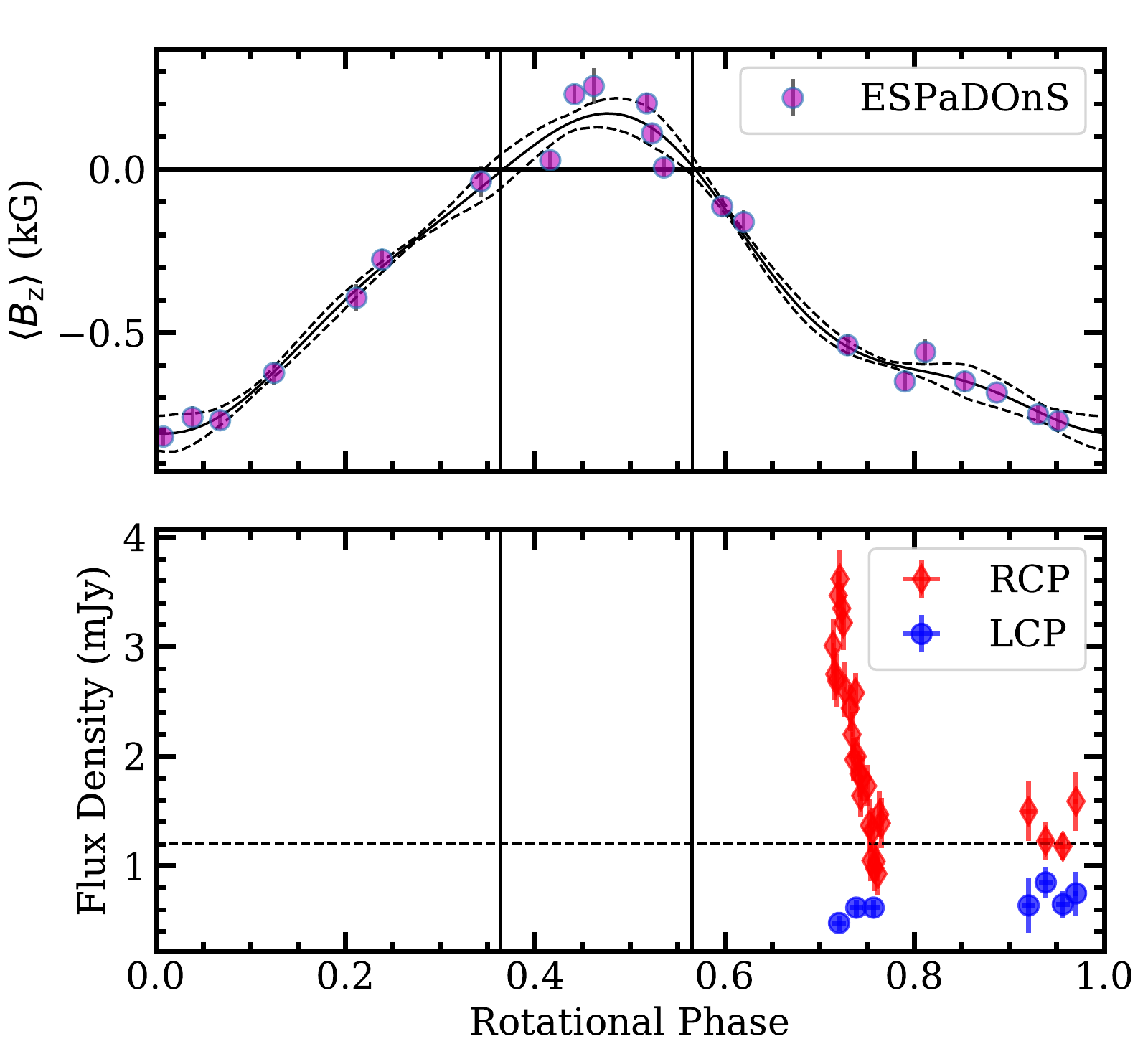}
    \caption{{\color{black}\textit{Top:} The variation of \bz~with rotational phase of HD\,61556 \citep{shultz2015}. The fitted curve corresponds to Eq.\ref{eq:bz_curve} with $n=3$. \textit{Bottom:}
     The light curves at 687 MHz. The smallest averaging time is 5 minutes, or 0.002 rotation cycles. The dashed horizontal line corresponds to the basal flux density listed in Table \ref{tab:star_properties}.}
     }
    \label{fig:hd61556}
\end{figure}

HD\,61556 is the star with the poorest rotational phase coverage in our sample. This star was found to emit highly circularly polarized ($\approx 76\%$) radio emission at 888 MHz by \citet{pritchard2021} using the Australian Square Kilometre Array Pathfinder (ASKAP) telescope, which suggests that the star is an MRP. This is further supported by its high $X$-factor \citep{das2022}. After considering the uncertainty in the $X$-factor, and the scatter of the empirical relation, its predicted peak flux density ranges between $\sim 3$ and $30$ mJy. 

Figure \ref{fig:hd61556} shows the radio light curves (bottom panel) along with the stellar \bz~variation (top panel); we detected a strong enhancement in RCP at phase $\approx 0.72$, which is offset from the nearest magnetic null by $\approx 0.16$ rotation cycles. {\color{black}The peak flux density is $3.6\pm 0.3$ mJy} with a circular polarization fraction of $\approx 76\%$. The lower limit to the brightness temperature is $\sim 10^{11}$ K, however the timescale of variation $\Delta\phi_\mathrm{rot}$ is only $\approx 0.04$ cycles. Based on these observations, we attribute the enhancements to ECME.

\subsection{HD\,182180}\label{subsec:hd182180}
\begin{figure}
    \centering
    \includegraphics[width=0.45\textwidth]{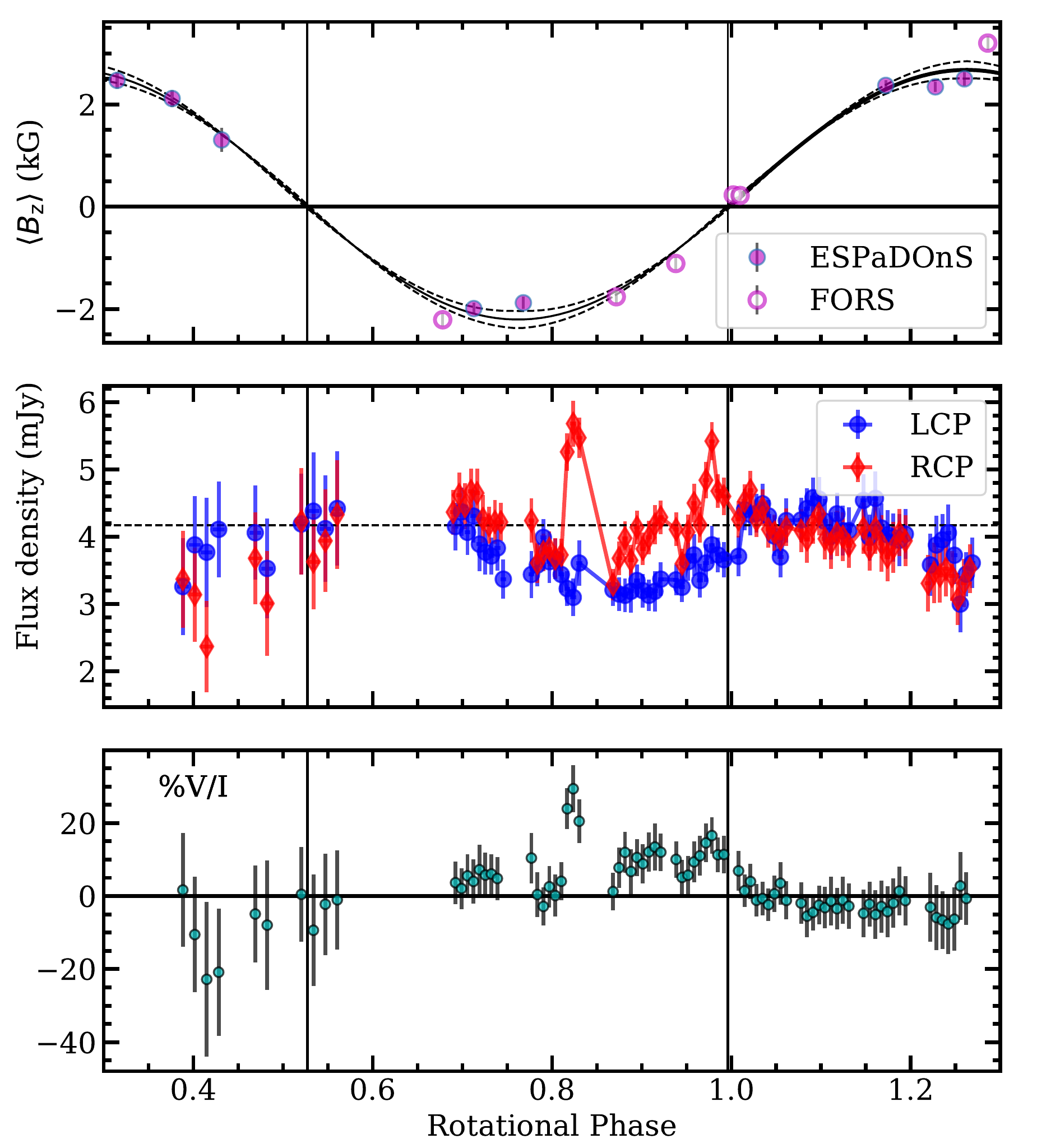}
    \caption{\textit{Top:} The \bz~curve of HD\,182180. The ESPaDOnS and FORS data were reported by \citet{oksala2010} and \citet{rivinius2010} respectively. A sinusoidal fit ($n=1$ in Eq.\ref{eq:bz_curve}) was performed to the data which gives the phases for the two magnetic nulls as $0.53\pm0.02$ and $0.00\pm0.02$ (marked with the vertical lines).
    \textit{Middle:}
    The light curves of HD\,182180 over 618--687 MHz. The smallest integration time is 5 minutes or 0.007 rotation cycle. The dashed horizontal line corresponds to the basal flux density listed in Table \ref{tab:star_properties}.
    \textit{Bottom:} The percentage circular polarization fraction that clearly shows the multiple enhancements observed in RCP near the magnetic null around phase 0.0.
    }
    \label{fig:hd182180_rr_ll}
\end{figure}

\begin{figure*}
    \centering
    \includegraphics[width=0.8\textwidth]{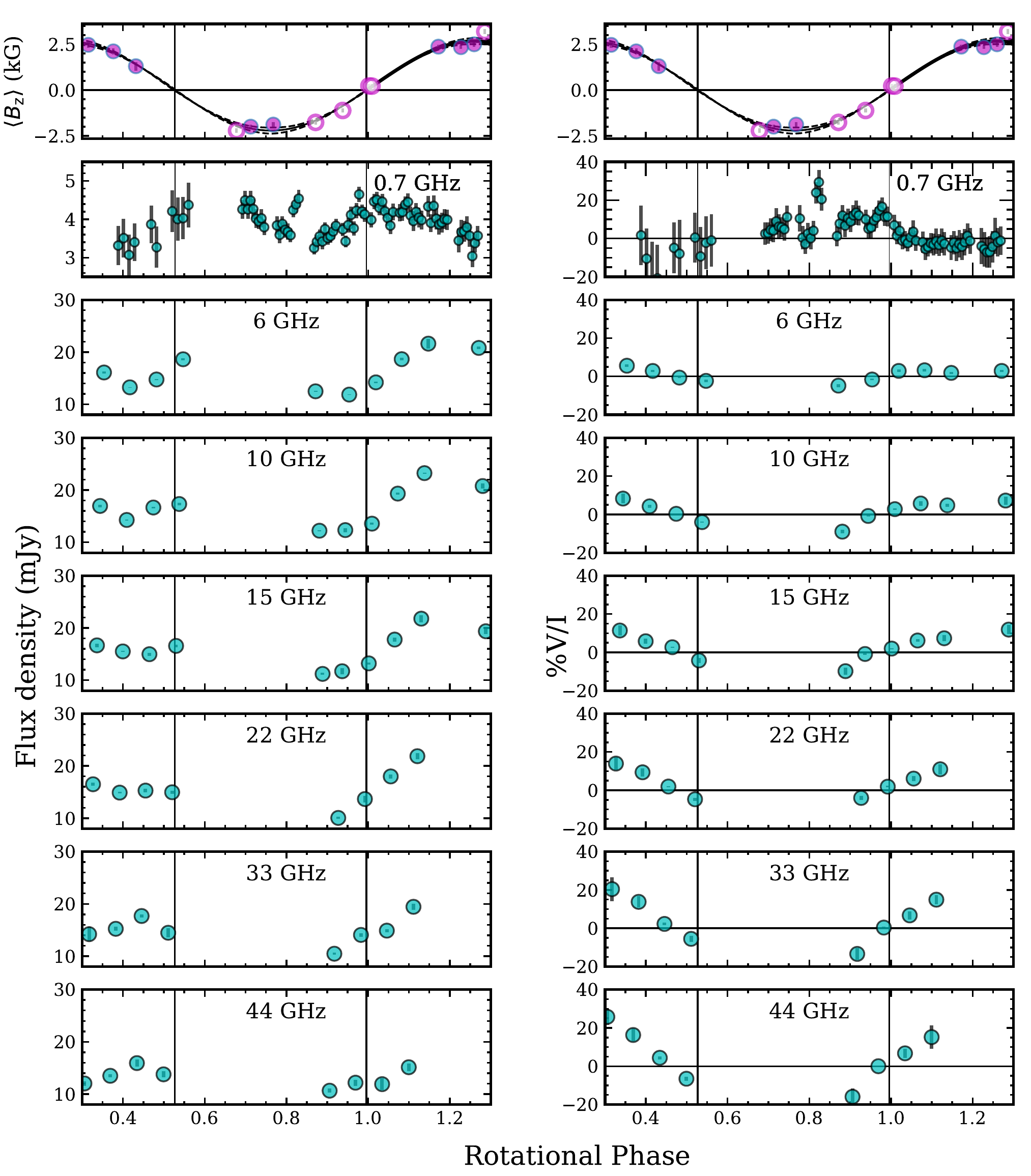}
    \caption{The light curves of HD\,182180 for total intensity (left) and fractional circular polarization (right) over 0.7--44 GHz. Also shown are the \bz~variation (top panel). The data at GHz frequencies were reported by \citet{leto2017}. The magnetic nulls are marked with the vertical lines.}
    \label{fig:hd182180_IV}
\end{figure*}
HD\,182180 was observed for one complete rotation cycle by \citet{leto2017} over the frequency range of 6--44 GHz. The star shows strong rotational modulation down to their lowest frequency of observation \citep[the amplitude of the flux density variation is 60\% relative to the median at 6 GHz,][]{leto2017}, but no sign of ECME was detected. A strong correlation between the rotational modulation of the radio flux density and that of \bz~was observed. Interestingly, a phase lag of $\approx 0.1$ cycles was found between one of the magnetic nulls and the minimum of the 6 GHz total intensity light curve \citep{leto2017}. Such a `phase lag' was also seen for $\rho\,\mathrm{Oph\,A}$ \citep{leto2020} and CU\,Vir
\citep{das2021}. Whereas in the former case, the discrepancy was attributed to the presence of non-dipolar components to the surface magnetic field \citep{leto2020}, the latter was considered a limitation of the adopted ephemeris \citep{das2021}.

The star was predicted to be an MRP at our observing frequency by \citet{das2022}. The $X$-factor of the star lies in the range 0.4--2.8 after considering the error bars in its $B_\mathrm{max}$ and $T_\mathrm{eff}$ (Table \ref{tab:star_properties}). The corresponding expected excess peak flux density is $\sim 1-10$ mJy.

Figure \ref{fig:hd182180_rr_ll} shows the light curves of the star obtained by combining our uGMRT data with existing archival GMRT data. The rotational modulation is evident even at sub-GHz frequencies. Superimposed on the rotationally modulated light curves, we see clear enhancement in RCP flux density, 
between phases 0.8 and 1.0.
The enhancement consists of three `sub-enhancements'.
The value of $\Delta\phi_\mathrm{rot}$ for each of these enhancements is $<0.05$.
The lower limit to the brightness temperature, assuming an emission site as large as the stellar disk, is greater than $10^{12}$ K. These characteristics confirm that indeed HD\,182180 produces ECME at our frequency of observation. 
{\color{black}The maximum observed flux density is $5.7\pm0.3$ mJy (RCP), so that the excess peak flux density is $\approx 2.6$ mJy (using the LCP flux density at the same phase), consistent with the predicted range.}

{\color {black}
In Figure \ref{fig:hd182180_IV}, we show the total intensity (left) and circular polarization (right) light curves for the star over the frequency range of 0.7--44 GHz \citep[The data at GHz frequencies were obtained from][]{leto2017}. Although the previous observations covered the magnetic null phases, they did not cover the rotational phase at which the strongest ECME enhancement at sub-GHz frequency was observed (phases 0.80--0.85). This raises the possibility of the presence of ECME even at GHz frequencies, which can be verified by re-observing the star at those frequencies.
}

The complex ECME pulse profile is similar to those observed for HD\,19832 and HD\,145501C \citep{das2022}. For these two stars, the enhancements were observed encompassing a magnetic null. In case of HD\,182180, the mid-phase between the two RCP enhancements is $\approx 0.9$, which  is offset from the nearest magnetic null by 0.1 cycles. This is remarkable as this is the phase lag observed between the gyrosynchrotron modulation and \bz~modulation of the star \citep{leto2012}. This suggests that a better strategy to find ECME is probably to look for flux density enhancement around the rotational phases corresponding to the minima of the incoherent radio light curve, rather than the nulls of \bz. 

Another interesting point to note here is that the sign of the circular polarization due to ECME at 0.7 GHz is opposite to that due to gyrosynchrotron at similar rotational phases at higher radio frequencies (Figure \ref{fig:hd182180_IV}). For the incoherent radio emission, the sign of the circular polarization varies in the same manner as that of \bz, indicating that the emission is in the extraordinary mode. In the case of ECME, it has been shown that it is not trivial to deduce the magneto-ionic mode based on the observed sign of the circular polarization \citep{das2020a}. Assuming that the ECME magneto-ionic mode is also extraordinary, the observation of RCP enhancement at rotational phases where \bz~is negative would imply strong propagation effects in the magnetosphere, which is consistent with the fact that the star has a large obliquity \citep[which is when the propagation effects become most important,][]{das2020a}. 
Alternately, our observations suggest that the amplified magneto-ionic mode of the ECME is the ordinary mode \citep[e.g.][]{leto2019}.

\section{Discussion}\label{sec:discussion}
The empirical relation between ECME luminosity and the $X$-factor (a function of stellar magnetic field and temperature) proposed by \citet{das2022} is the first of its kind that provides us with clues to understand the onset of coherent radio emission in hot magnetic stars. In its current stage, the relation should be treated more as a tool to assess if a star is likely to be an MRP or not, rather than obtaining an estimate of the expected peak ECME flux density. This is because there is considerable scatter surrounding the relation, such that the uncertainty associated with the predicted flux densities could be $\gtrsim$1 dex. The only way to improve/revise the relation
is to increase the sample size, i.e. by discovering more MRPs. 


In the following subsections, we discuss our main results.

\subsection{ECME from HD\,35502}\label{subsec:no_ecme_hd35502}
\begin{figure}
    \centering
    \includegraphics[width=0.45\textwidth]{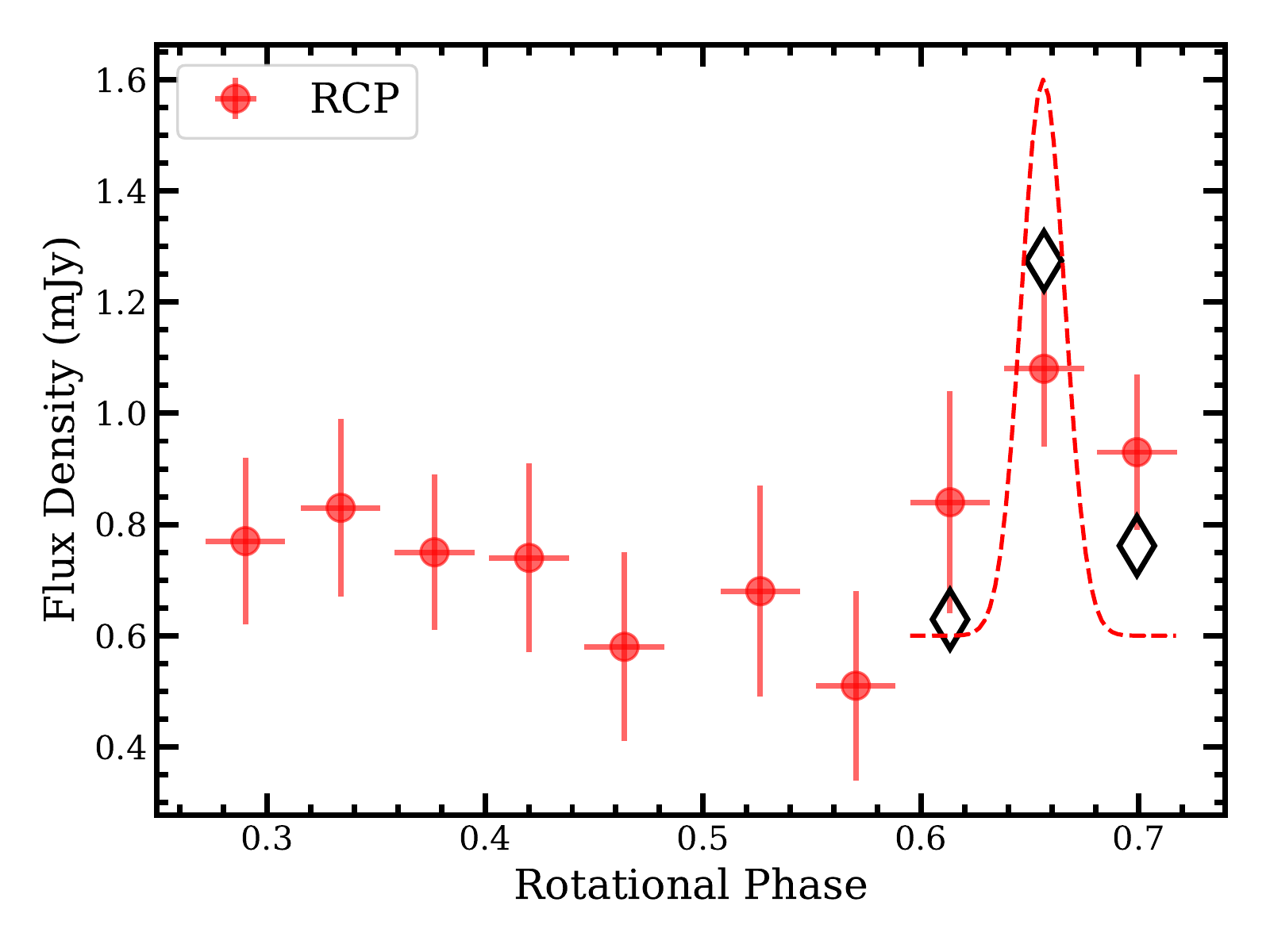}
    \caption{Figure illustrating the sampling effect in detecting an ECME pulse. The circles represent the observed RCP flux density from HD\,35502. The averaging time is 45 minutes or 0.04 rotation cycles. The red dashed curve shows a possible pulse, assumed to have a Gaussian profile with $\sigma=0.01$ rotation cycles. The integration time corresponding to the dashed curve is $\approx 0.002$ rotation cycles. The black unfilled diamonds show the effect of sampling this curve with a resolution of 0.04 rotation cycles. As can be seen, the signature of the pulse is significantly weakened due to poor time-resolution.}
   \label{fig:hd35502_hidden_pulse}
\end{figure}

This is the only case for which we could not confirm the prediction of the empirical relation. We observed a marginal enhancement in the RCP flux density over a relatively narrow rotational phase range (bottom panel of Figure \ref{fig:hd35502}), but our measurements are not precise enough to conclusively attribute the enhancement to ECME. In this subsection we discuss the situations under which the ECME produced by the star may not be observable.

Although we spanned a rotational phase window of width 0.4 cycles, we centred it at phase 0.5 \citep[the magnetic null according to][]{sikora2016}, whereas \citet{shultz2018} found the magnetic nulls to lie at around phases 0.4 and 0.6. Our observation covers 0.3--0.7 rotation cycles, however, an offset as large as 0.15 cycles was observed between the peak of a sub-GHz ECME pulse of CU\,Vir and the nearest magnetic null \citep{das2021}. Thus, it is possible that the rotational phases corresponding to ECME enhancement lies outside the observed rotational phase window.


There is, however, a more acute problem associated with this star in that it barely satisfies the geometrical condition for observation of ECME, which is that the sum of the inclination angle ($i$) and obliquity ($\beta$) should be greater than or equal to $90^\circ$. For this star, $i+\beta=96^\circ\pm 2^\circ$ (Table \ref{tab:star_properties}), thus the condition is only marginally satisfied. 
The failure to detect ECME from HD\,35502 constrains the geometrical condition of the maser. In fact, if the (net) maser emission direction w.r.t. the magnetic axis is lower than $80^\circ$, 
the beam of the maser will be never aligned with the line of sight within the range of phases covered by the reported observations.
It is hence possible that the ECME beam never aligns with the line of sight, which will either diminish the magnitude of the enhancement, or will lead to no enhancement at all.

If the geometric condition is indeed responsible for suppressing the observable ECME from HD\,35502, one would expect a similar effect for HD\,61556 also for which $i+\beta=94^\circ \pm 13^\circ$, unless for the latter, the actual values of the inclination angle and the obliquity are closer to their respective maximum allowed values (i.e. $i\approx 42^\circ$ and $\beta\approx 65^\circ$). Between HD\,35502 and HD\,61556, the former has a stronger magnetic field and shorter rotation period, but the two have very similar radii and effective temperatures (Table \ref{tab:star_properties}). This makes HD\,35502 a more favorable star to produce non-thermal radio emission \citep{leto2021,shultz2022,das2022}. Despite that, the expected peak ECME flux density from HD\,35502 is significantly lower than that for HD\,61556 due to the relatively large distance of the former. This effect (small expected flux density) can alone be responsible for not detecting a clear ECME signature from HD\,35502. From Figure \ref{fig:hd35502}, we can see that the averaging time corresponding to each flux density measurement of HD\,35502 is 45 minutes or 0.04 rotation cycles, whereas for HD\,61556, it is only 5 minutes or 0.002 rotation cycles. So far, it is not clear what determines the width of an ECME pulse. At sub-GHz frequency, ECME pulse width (FWHM) was found to vary from $\approx 0.02$ rotation cycles \citep[HD\,12447,][]{das2022} to $\sim 0.1$ \citep[CU\,Vir,][]{das2021}. If we assume that HD\,35502 produces ECME, which is observable over phases 0.60 to 0.72 with a peak at phase 0.66 (the phase at which maximum flux density was observed), and has a gaussian profile with $\sigma=0.01$ rotation cycles (FWHM$\approx$0.02), we find that averaging over 0.04 rotation cycle will significantly weaken the pulse-signature to the extent shown by our observations (Figure \ref{fig:hd35502_hidden_pulse}). Thus, a very likely reason for not detecting an ECME pulse from HD\,35502, could simply be the inadequate sensitivity of our measurements.

Thus with the current data, we are unable to confirm whether the star is an MRP or not. More sensitive observations covering a wider rotational phase range of the star, and/or observations performed at other frequencies, that could be likely radiated within a radiation beam with a higher opening angle (w.r.t. the magnetic axis), will be necessary to distinguish between these possibilities.

\subsection{The obliquity effect on ECME}\label{subsec:obliquity}
\citet{das2022} proposed that an obliquity close to $90^\circ$ might lead to a multi-peaked ECME pulse profile at the same circular polarization due to propagation effects. This was based on their observation of such pulse profiles from three of their sample stars, all of which have obliquities close to $90^\circ$, and also from HD\,142990, which also has an obliquity close to 90$^\circ$ \citep{das2019a}. They explained this phenomenon to be a result of refraction suffered by emission propagating in a magnetosphere with a highly complex plasma distribution, which in turn is a consequence of the large obliquity \citep{townsend2005}. In our sample of stars, HD\,182180 is the only star with an obliquity close to $90^\circ$ (Table \ref{tab:star_properties}), and interestingly, we did observe a multi-peaked RCP pulse from this star (Figure \ref{fig:hd182180_rr_ll}). This provides strong support to the notion that ECME pulse profiles are greatly affected by the stellar magnetospheric plasma \citep{das2020b}, and that complex pulse profiles are associated with large obliquities \citep{das2022}.

It is worth noting that \citet{shultz2022} observed a negative correlation between incoherent gyrosynchrotron luminosity and the obliquity, which they explained to be a consequence of reduced amount of confined plasma in the magnetosphere with increasing obliquities \citep[as predicted by the `Rigidly Rotating Magnetosphere' or RRM model,][]{townsend2005}. We, however, do not find any correlation between obliquity and \Le~with the current sample of MRPs. 

\subsection{The importance of magnetic field mapping in ECME characterization}\label{subsec:magnetic_map}
The observed ECME characteristics are very often different from those expected from a star with an axi-symmetric dipolar magnetic field. Although, such discrepancies have been attributed to a more complex magnetic field topology than that of a dipole \citep[e.g.][]{leto2019,das2019a}, \citet{das2022} showed that such a one-to-one correlation between non-ideal ECME characteristics and complexity of the surface magnetic field, is absent in the sample of MRPs, and other effects, such as obliquity and precision of the adopted ephemeris should be taken into account (in addition to the magnetic field) to explain the discrepancies. It is, however, to be noted that only a small number of stars have available surface magnetic field maps \citep[e.g.][etc.]{kochukhov2014,kochukhov2017,oksala2015}, and for the rest, the deviation of the \bz~variation with rotational phase from a sinusoidal variation is taken as a proxy to indicate the complexity of the surface magnetic field. 
Since \bz, by definition, is the line-of-sight magnetic field averaged over the stellar surface, it is only weakly sensitive to departures from a centered dipole. In addition, \bz~is also affected by non-magnetic photospheric phenomena, such as the non-uniform distribution of chemical abundances \citep[e.g.][]{yakunin2015}.
Magnetic field maps, on the other hand, are more useful in predicting how the field will vary as a function of height from the stellar surface, which is likely to play an important role in the beaming effect of the ECME pulses at different frequencies \citep[e.g.][]{melrose1982,trigilio2011}. For low mass stars, \citet{llama2018} have used magnetic maps through Zeeman Doppler Imaging \citep[ZDI,][]{semel1989} to predict coronal radio and X-ray emission properties. The difference between the surface magnetic field and that at the height of radio emission is also important to understand the phase lags that are sometimes observed between the rotational modulation of \bz~and incoherent radio emission \citep{leto2017}. With the magnetic field well-characterized, it will also be easier to understand the relative importance of other effects suggested by \citet{das2022}.

Thus to characterize the ECME phenomenon, acquiring spectropolarimetric observations covering full rotation cycles of the stars will be as important as acquiring more radio observations, the importance of which is described in the subsequent subsections.

\subsection{The scaling relation between ECME luminosity vs stellar parameters following the addition of new MRPs}\label{subsec:new_X_factor}
\begin{figure*}
    \centering
    \includegraphics[width=0.95\textwidth]{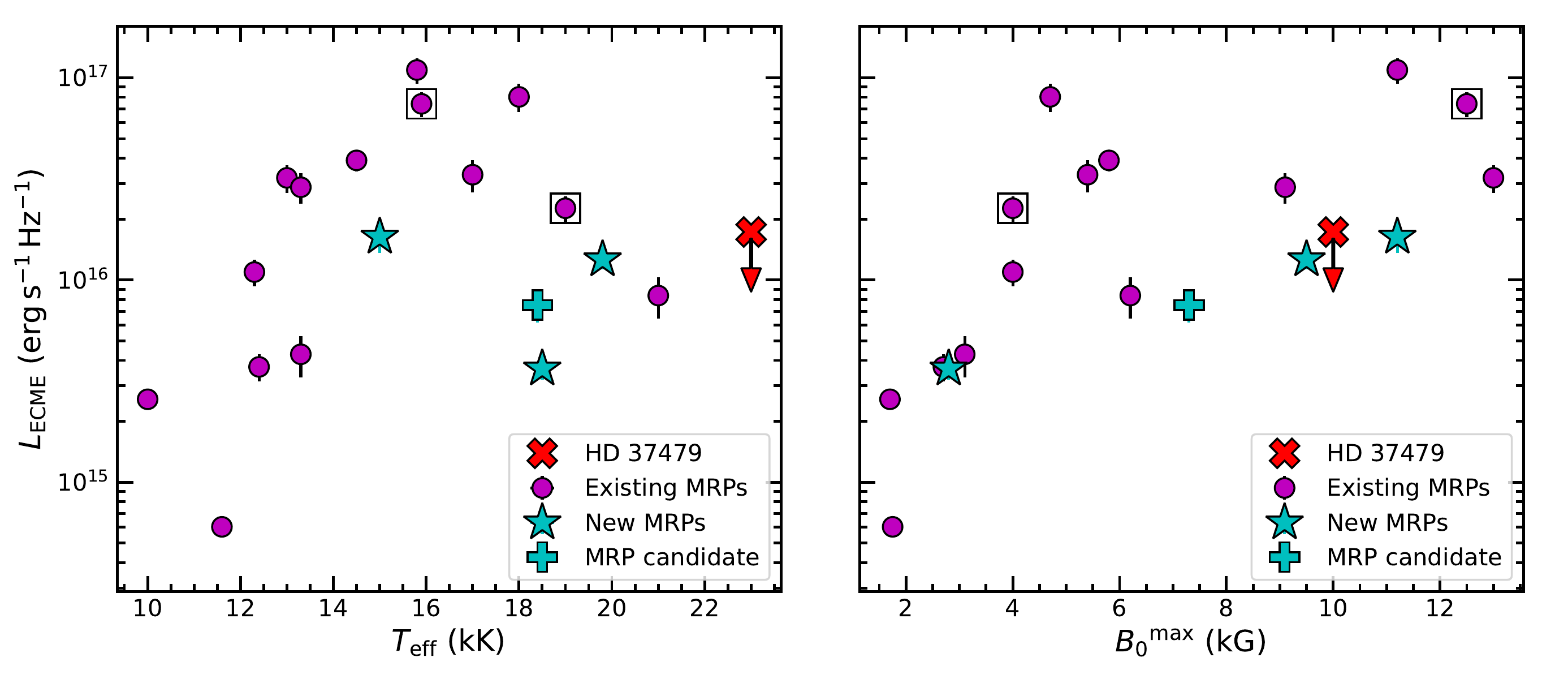}
    \caption{Excess peak ECME flux density vs stellar effective temperature $T_\mathrm{eff}$ (left), and the maximum surface magnetic field strength $B_0^\mathrm{max}$ (right). The three new confirmed MRPs (HD\,36526, HD\,61556 and HD\,182180) are marked with cyan stars. The MRP candidate HD\,35502 is marked with a cyan `+' symbol. The non-detection of ECME for HD\,37479 is shown by the downward arrow (marker `x'). The markers surrounded by squares represent the stars HD\,142301 and HD\,147933, for which reported sub-GHz measurements of ECME flux density do not exist \citep{leto2019,leto2020}. 
    For these two stars, we have used their flux density measurements at 1.5 GHz \citep[HD\,142301,][]{leto2019} and 2.1 GHz \citep[HD\,147933,][]{leto2020}.
    The remaining data were reported by \citet{das2018,das2019b,das2019a,das2022}.}
    \label{fig:ecme_vs_T_B}
\end{figure*}

A key motivation behind our radio observation was to refine the empirical relation of \citet{das2022}. 
We first examine the correlation of the excess peak ECME luminosity \Le~with $T_\mathrm{eff}$ and $B_0^\mathrm{max}$ (Figure \ref{fig:ecme_vs_T_B}). 
It is to be noted that for HD\,36526 and HD\,64740, the peak flux densities are probably underestimated. For the former, we had to average over 45 minutes (0.02 rotation cycles) in order to attain sufficient sensitivity to detect its radio emission, whereas for the latter, we could not observe the complete pulse. This was also the case for HD\,176582 reported by \citet{das2022}. This effect is however unlikely to be important at this stage since the scatter around the correlation under consideration is $\sim 1$ dex.
We find that after the addition of the three new data points (HD\,36526, HD\,61556 and HD\,182180; HD\,35502 is not included in the analysis as its status as an MRP is yet to be confirmed), the parabolic dependence on $T_\mathrm{eff}$ remains unaffected. The Spearman's rank correlation coefficient between \Le~and the quantity ${(T_\mathrm{eff}-16.5)}^2$ (The temperature dependent part of the $X$-factor) comes out to be $-0.7$ with a p-value of 0.004. With magnetic field strength also, we obtained a correlation co-efficient of similar magnitude ($+0.7$), and a p-value of 0.003.



\begin{figure}
    \centering
    \includegraphics[width=0.45\textwidth]{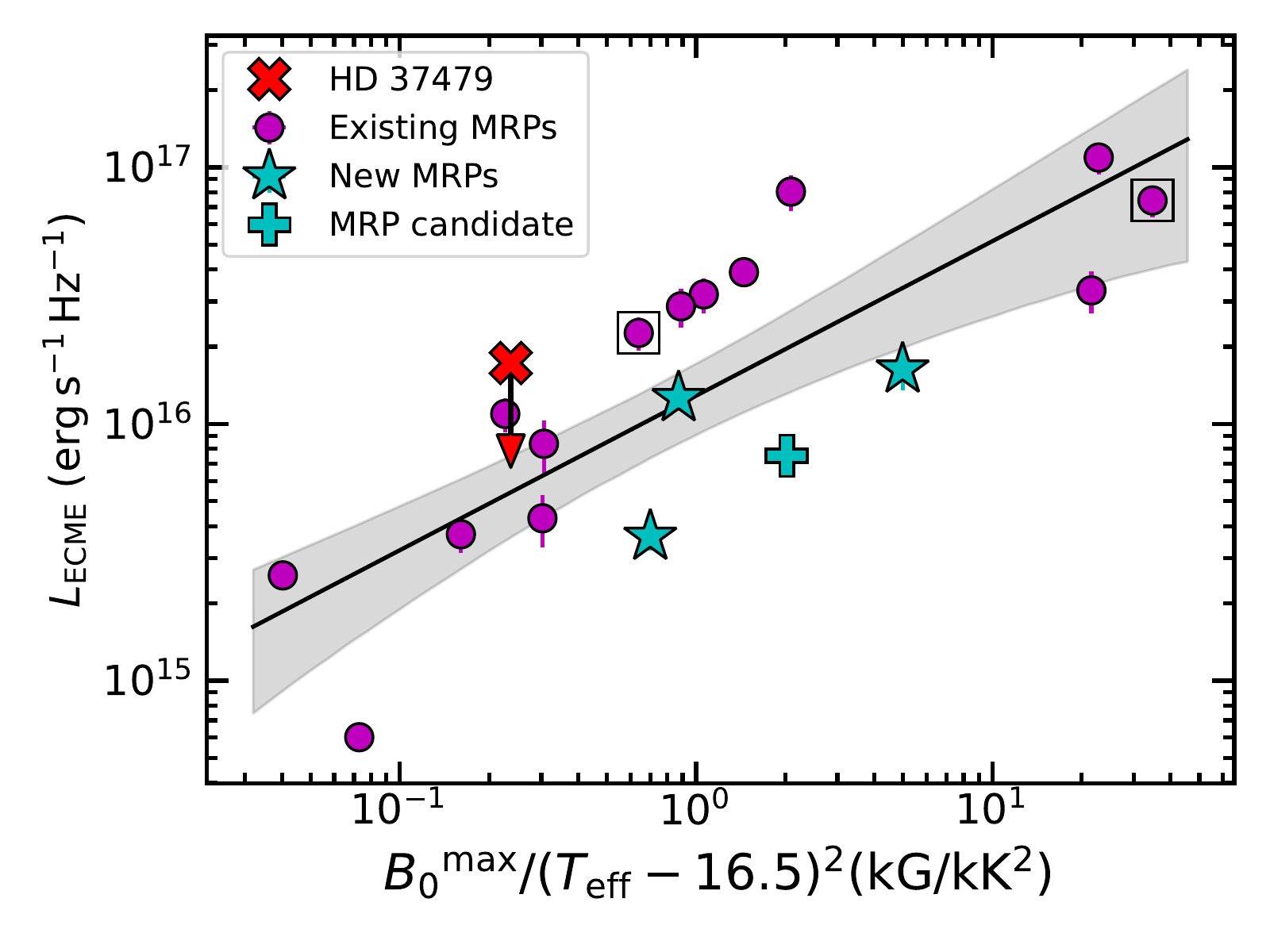}
    \caption{The correlation between the peak ECME flux density (at 0.7 GHz) vs the $X$-factor of \citet{das2022}. The significance of different markers are given in the caption of Figure \ref{fig:ecme_vs_T_B}.}
    \label{fig:ecme_X_factor}
\end{figure}

We now examine the effect of the addition of the three new MRPs on the empirical relation of \citet{das2022}. Figure \ref{fig:ecme_X_factor} shows the relation between peak ECME luminosity and the $X$-factor. The best-fit relation comes out to be \Le$\propto X^{0.6\pm0.1}$, consistent with the relation obtained by \citet{das2022}: \Le$\propto X^{0.8\pm0.1}$. The corresponding Spearman's rank correlation co-efficient is $+0.8$ with a p-value of $10^{-5}$. Thus the existing empirical relation seems adequate for the current data.

To summarize, we find that the basic relations obtained by \citet{das2022} between peak ECME luminosity and stellar parameters remain valid following the addition of new MRPs. 
The emerging relation is in stark contrast with that obtained for the incoherent radio luminosity, in which the temperature does not play a role at all, and a strong correlation was obtained with the magnetic flux \citep[a function of the polar magnetic field strength and the stellar radius,][]{leto2021,shultz2022}. The next section (\S\ref{subsec:gyro_ecme}) provides a detailed discussion on this aspect.


\subsection{Are incoherent and coherent radio emission powered by the same phenomenon?}\label{subsec:gyro_ecme}
\begin{figure*}
    \centering
    \includegraphics[width=0.45\textwidth]{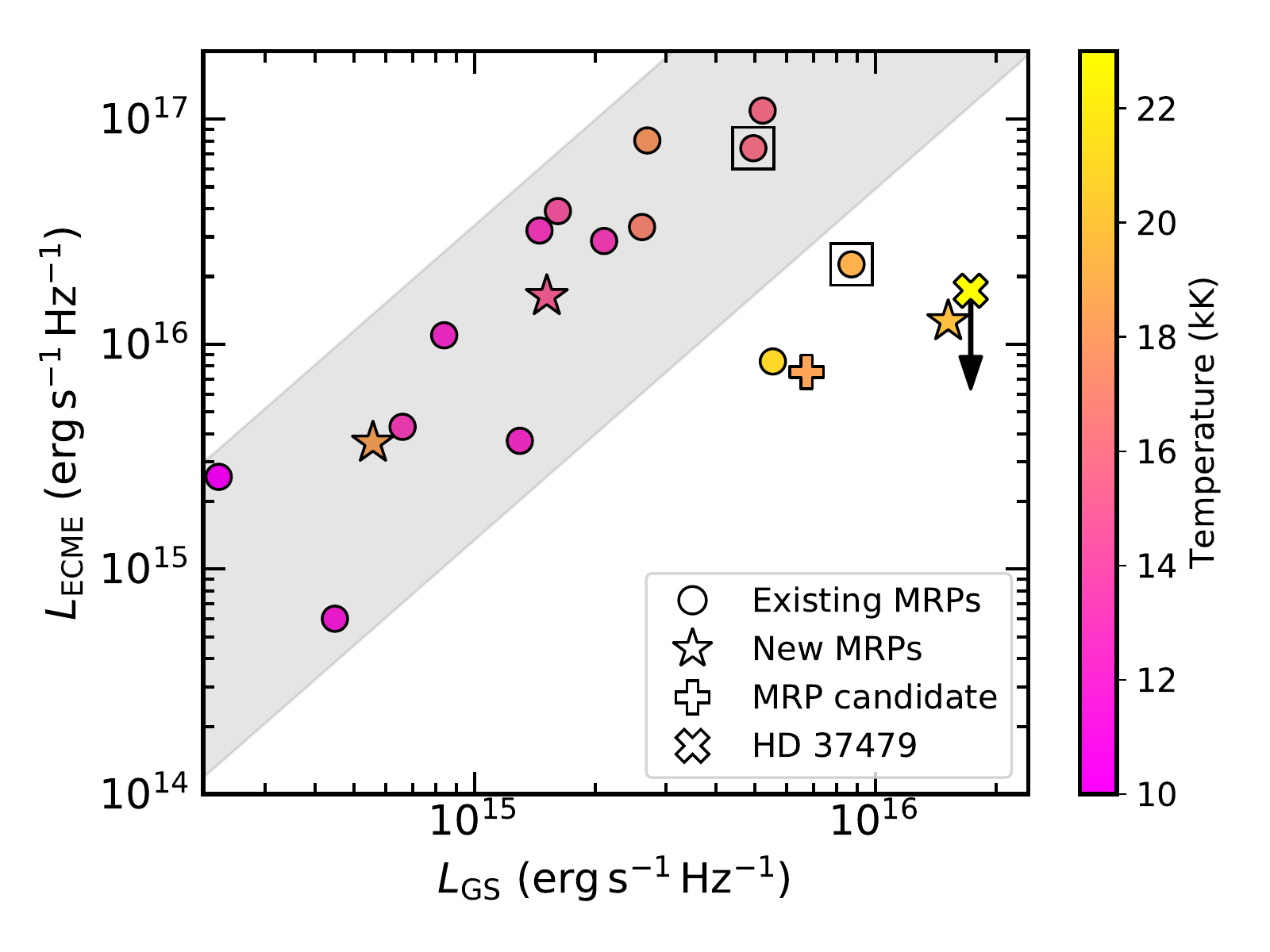}
    \includegraphics[width=0.45\textwidth]{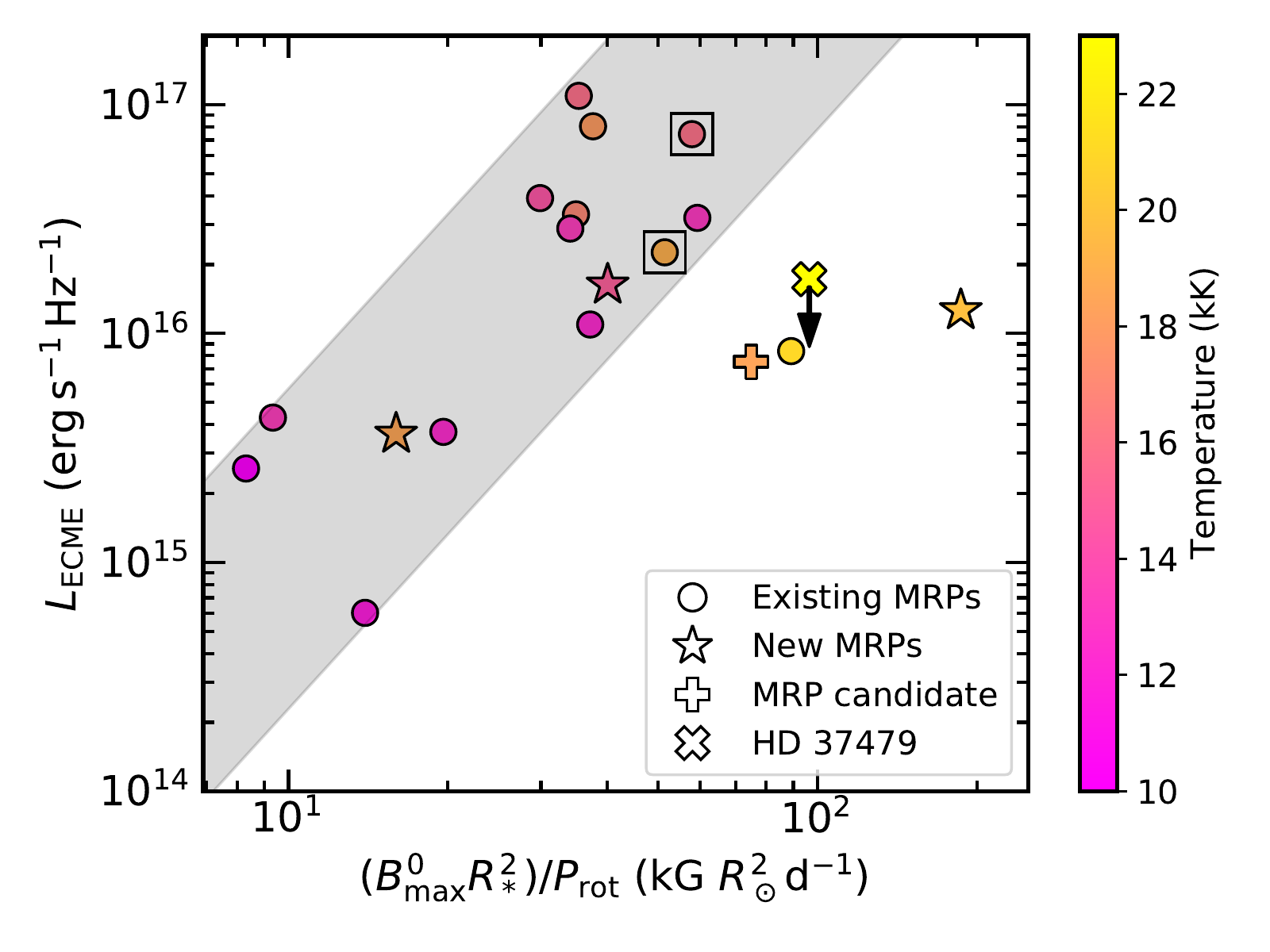}
    \caption{\textit{Left:} The comparison of gyrosynchrotron flux density and peak ECME flux density (in excess of the gyrosynchrotron flux density). \textit{Right:} The excess ECME luminosity vs the quantity $B^0_\mathrm{max}R_*^2/P_\mathrm{rot}$. The latter was proposed to control the incoherent radio luminosity from ordered magnetospheres by \citet{leto2021}, also confirmed by \citet{shultz2022,owocki2022}. The upper limit represents HD\,37479. The marker colors are in accordance with the stellar temperature. The quantities (along the X and Y axes) are correlated except for the top four hottest stars in the sample. The significance of different markers are the same as described in the caption for Figure \ref{fig:ecme_vs_T_B}.}
    \label{fig:gyro_ecme}
\end{figure*}

Recently, the idea of generation of non-thermal electrons in hot stars' magnetospheres have been revolutionized by the work of \citet{leto2021} and \citet{shultz2022}. They surprisingly discovered that the incoherent radio luminosity is indifferent to the mass-loss rate (or proxies such as effective temperature or bolometric luminosity), but has an inverse dependence on the stellar rotation period, and a positive dependence on the unsigned magnetic flux (a function of the dipole strength and stellar radius). \citet{owocki2022} provided a theoretical framework to explain this observation, according to which the gyrosynchrotron emission is powered by magnetic reconnection triggered by centrifugal breakout (CBO) events. The CBO phenomenon has a strong dependence on the stellar rotation period 
which explains the dependence of the incoherent radio luminosity on the stellar rotation period. This also provides a natural explanation for the observed correlation between magnetospheric $\mathrm{H\alpha}$ emission and incoherent radio emission \citep{shultz2022}, since the former is also connected to CBO \citep{shultz2020,owocki2020}. Thus the CBO framework provides a unified model for the two kinds of magnetospheric emission.
Interestingly, \citet{leto2021} speculated on the existence of two channels for production of non-thermal electrons: a radiation belt located inside the `closed' magnetosphere that supplies the non-thermal electrons emitting the gyrosynchrotron emission, and an equatorial magneto-disk located outside the closed magnetosphere (where magnetic reconnection takes place) that supplies the non-thermal electrons emitting ECME.

From \S\ref{subsec:new_X_factor}, we already find an important difference between the dependence of incoherent and coherent radio luminosity on stellar parameters, which is the involvement of temperature and no role of rotation period in case of coherent emission  \citep[correlation with the rotation period was examined by][but with null result]{das2022}. To inspect a possible connection between the incoherent and coherent radio emission, we plot the peak (excess) luminosity due to ECME against the basal gyrosynchrotron luminosity $L_\mathrm{\textsc{gs}}$ (taken as $S_\mathrm{\textsc{gs}}\times d^2$, where $S_\mathrm{\textsc{gs}}$ is the basal gyrosynchrotron flux density) in the left of Figure \ref{fig:gyro_ecme}. Excluding HD\,142301 and HD\,147933 (we have excluded HD\,147932 throughout), all the data points correspond to frequencies 0.6--0.7 GHz. The markers are color-coded in accordance with the respective temperature. 
It is clear that, for the cooler confirmed MRPs (i.e. HD\,35502 is not considered), there is a tight correlation between gyrosynchrotron luminosity and the ECME luminosity. The Spearman rank correlation co-efficient between the two quantities, excluding the hottest stars in our sample (equivalent to considering stars with $T_\mathrm{eff}<19\,$ kK), is $+0.9$ with a p-value of $10^{-7}$. The temperature dependence in the case of ECME luminosity is reflected as the loss of correlation with the incoherent radio luminosity for the hotter stars.

The very strong correlation observed between incoherent and coherent radio luminosity for the relatively cooler stars in our sample motivated us to examine the correlation between the ECME luminosity and $\tilde{L}_\mathrm{\textsc{cbo}}$, defined as:
\begin{align}
    \tilde{L}_\mathrm{\textsc{cbo}}&=\frac{B^0_\mathrm{max}R_*^2}{P_\mathrm{rot}} \label{eq:l_cbo}
\end{align}
{\color{black}Note that this quantity does not have a dimension of luminosity, but is proportional to the CBO luminosity $L_\textsc{cbo}$ derived by \citet{owocki2022} for their split monopole case.}
$L_\mathrm{\textsc{cbo}}$ has been
found to be responsible for controlling the incoherent radio luminosity from a wide varieties of ordered magnetospheres \citep{leto2021,shultz2022}, and is consistent with the scenario in which the radio emission is powered by CBO events \citep{shultz2022,owocki2022}. 
As expected, we observe a correlation between coherent radio luminosity and $\tilde{L}_\mathrm{\textsc{cbo}}$ except for the hottest stars in the sample (right of Figure \ref{fig:gyro_ecme}). The corresponding correlation co-efficient is however smaller in this case (0.7). Note that excluding the stars with $T_\mathrm{eff}\geq 19$ kK, the correlation co-efficient of coherent radio luminosity with magnetic field strength alone, increases to 0.8 with a p-value of 0.001. Thus inclusion of stellar radius and rotation period actually deteriorates the correlation. In other words, with the current MRP sample, without the information about the dependence of incoherent radio luminosity on stellar parameters \citep[which is a much more robust relation owing to the sample size,][]{leto2021,shultz2022}, and the observed strong correlation between incoherent and coherent radio luminosity (for $T_\mathrm{eff}<19$ kK), it is not possible to infer the dependence of the latter on stellar radius and rotation period.

\begin{figure}
    \centering
    \includegraphics[width=0.45\textwidth]{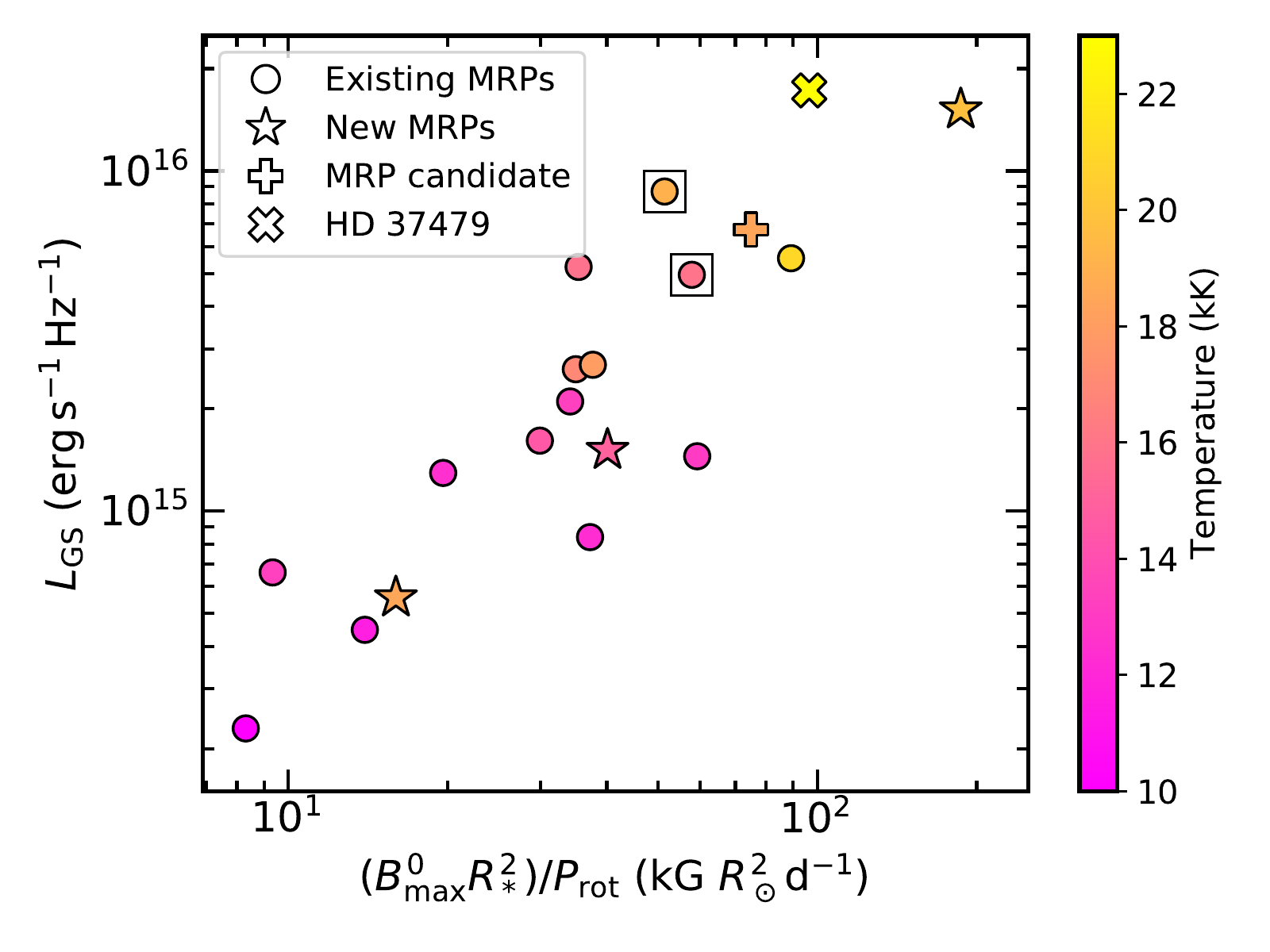}
    \caption{The correlation between incoherent radio luminosity and the ratio of magnetic flux to rotation period. 
    The significance of different markers are the same as described in the caption for Figure \ref{fig:ecme_vs_T_B}.
    }
    \label{fig:gyro_x_gyro_lum}
\end{figure}
The correlation observed for the relatively cool stars suggest that the incoherent and coherent radio emission are driven by the same phenomenon. The dependence on temperature in the case of ECME could be due to several reasons. As noted by \citet{leto2012}, for an O-star with a mass-loss rate of $10^{-6}\,\mathrm{M}_\odot/\mathrm{yr}$ and magnetic field of 1 kG, the necessary condition for ECME, which is that the plasma frequency ($\nu_\mathrm{p}$) should be smaller than the electron gyrofrequency ($\nu_\mathrm{B}$), may not be satisfied, which will inhibit the ECME generation process. The increasing difficulty in satisfying this condition with increasing temperature was also proposed by \citet{das2022} as a possible reason behind the observed dependence of ECME luminosity on temperature. There is also a possibility of higher absorption in the case of stars with higher temperature. This is, however, expected to affect both incoherent and coherent parts of the magnetospheric radio emission. But as the two types of emission (at the same frequency) arise in different parts of the magnetosphere (ECME is produced much closer to the star), the critical temperature beyond which the role of $T_\mathrm{eff}$ becomes significant, might be different for the two kinds of emission. This is supported by the fact that the hottest radio-bright star, HD\,64740 \citep[$T_\mathrm{eff}\approx 24$ kK,][]{shultz2019b}, was found to be underluminous w.r.t. the gyrosynchrotron scaling relation \citep{shultz2022}.
Note that we find a tight correlation between the incoherent radio luminosity and $L_\mathrm{\textsc{cbo}}$, including the hottest stars in the sample, and also HD\,35502 (Figure \ref{fig:gyro_x_gyro_lum}). 
Comparing Figure \ref{fig:gyro_x_gyro_lum} with the right of Figure \ref{fig:gyro_ecme}, it appears that the relation between the coherent radio luminosity and $\tilde{L}_\textsc{cbo}$ (for $T_\mathrm{eff}<19$ kK) is steeper than that with incoherent radio luminosity.

\begin{figure}
    \centering
    \includegraphics[width=0.45\textwidth]{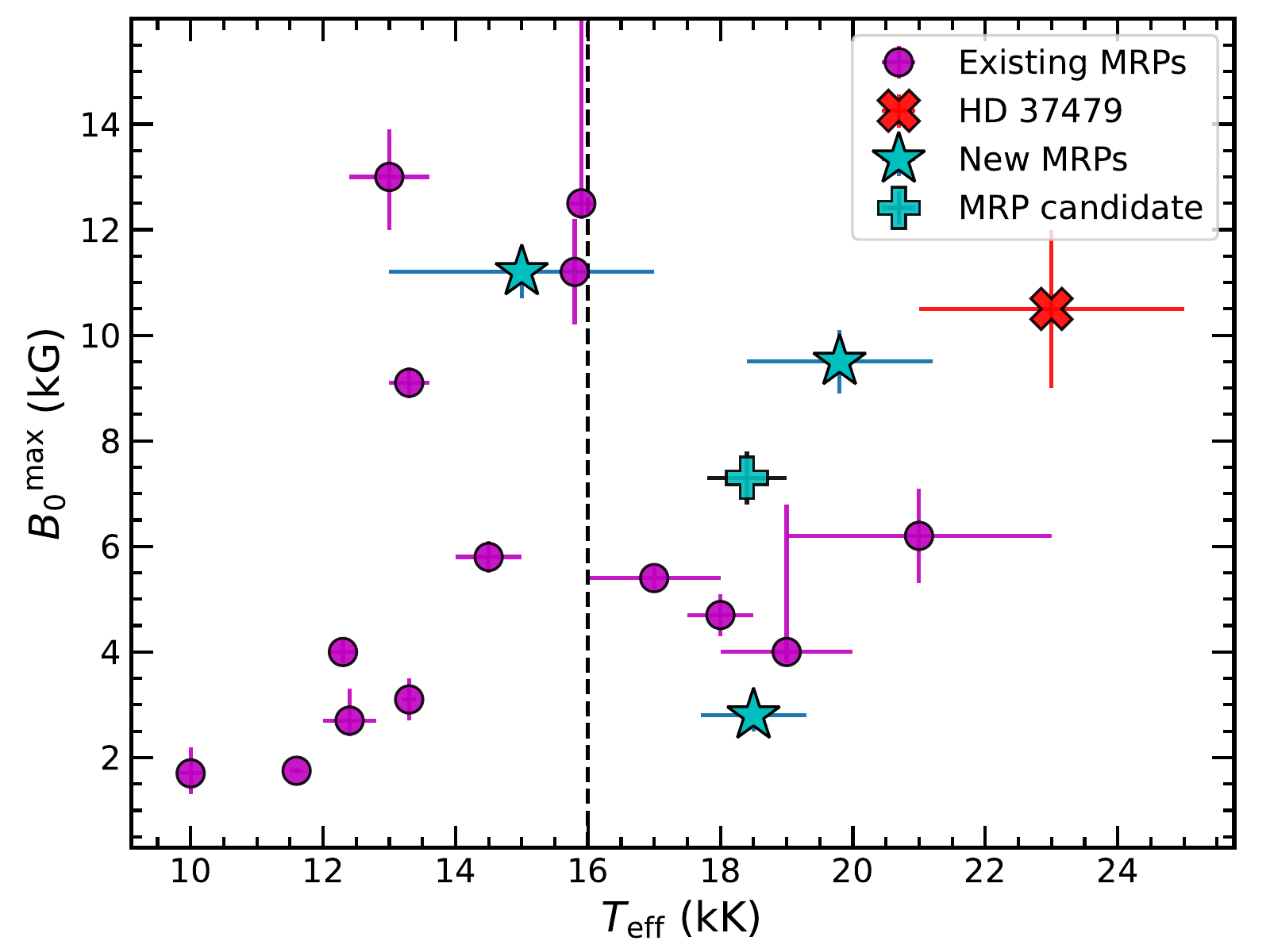}
    \caption{The relation between magnetic field strength and stellar effective temperature for the sample of MRPs. The new MRPs are marked with filled cyan `stars', and the MRP candidate HD\,35502 is marked with an unfilled `star'. The vertical dashed lines mark $T_\mathrm{eff}=16$ kK. Below this temperature, the two quantities are strongly correlated.}
    \label{fig:B_T}
\end{figure}

If the above scenario is correct, i.e., the same physical process drives both incoherent and coherent radio emission, and that the stellar temperature becomes important in ECME viability only above a certain temperature ($\approx 19$ kK), the observed positive correlation of ECME luminosity with $T_\mathrm{eff}$ up to around 16 kK (left of Figure \ref{fig:ecme_vs_T_B}) must arise due to an underlying correlation between $T_\mathrm{eff}$ and $B^0_\mathrm{max}$ in our sample. Indeed we found that for the MRPs with $T_\mathrm{eff}<=16$ kK, the magnetic field strength and the effective temperature are correlated with a Spearman rank correlation co-efficient of 0.7 and a p-value of 0.008 (see Figure \ref{fig:B_T}).

Based on these analysis, we suggest that both incoherent (gyrosynchrotron) and coherent (ECME) radio emission are powered by the same magnetospheric phenomenon, namely the centrifugal breakout of confined plasma from the magnetosphere. The stellar effective temperature does not play a role in driving ECME up to a `critical temperature' which lies around 19 kK. Above that temperature, the magnetosphere probably becomes dense enough such that the ratio between the plasma frequency and the electron gyrofrequency starts to become an important factor in the generation of ECME, or to make absorption a dominant effect, or both. Thus the $X$-factor of the form obtained by \citet{das2022}, which contains both magnetic field and temperature, is relevant only for stars with $T_\mathrm{eff}\gtrsim$ 19 kK. 
Below that temperature, the incoherent gyrosynchrotron radio luminosity (or $L_\mathrm{\textsc{cbo}}$) is itself a good indicator of the likelihood of a star to become an MRP.

If the reason for the role of $T_\mathrm{eff}$ becoming important for the hotter stars is the increasing value of the ratio $\nu_\mathrm{p}/\nu_\mathrm{B}$, the $X$-factor (for the hotter stars) is likely to be of the form ${(B_0^\mathrm{max})}^m/T_\mathrm{eff}^n;\,m>0,\,n>0$. On the other hand, if the reason is higher absorption (thus, not related to the ECME generation process), a stronger magnetic field will not be able to compensate for the increasing temperature, and the $X$-factor will probably be independent of the magnetic field.

{\color{black} 
Finally, we would like to point out a potentially important coincidence observed in Figure \ref{fig:gyro_ecme}, which is that all the stars with $T_\mathrm{eff}\gtrsim 19$ kK lie towards the right of the plots, such that one can also define the outliers based on a critical value of the gyrosynchrotron flux density or $L_\mathrm{\textsc{cbo}}$. In other words, correlation between $L_\mathrm{\textsc{ecme}}$ and $L_\mathrm{\textsc{cbo}}$ appears to exist below a certain value of the latter. At present, we do not have a physical explanation for this possibility, but this point will be worth checking in the future with the help of a larger sample of MRPs.
}

In the future, it will be important to acquire observations of more magnetic stars with $T_\mathrm{eff}\gtrsim 19$ kK so as to be able to infer the correct form of the $X$-factor for producing ECME by the early-B and O-type stars, and to
better understand the difference(s) in physical processes involved in the generation and/or propagation of incoherent and coherent radio emission. In particular, it will be extremely interesting to understand the significance of $T_\mathrm{eff}\approx 19$ kK. 


\section{Summary}\label{sec:summary}
The primary findings of this paper are summarized below:
\begin{enumerate}
    \item Discovery of three new MRPs: We discovered that HD\,36526, HD\,61556 and HD\,182180 produce ECME at sub-GHz frequencies. With the addition of these stars, the current number of confirmed MRPs has increased to 17. 
    \item Indication of the first confirmed non-MRP down to sub-GHz frequencies: By observing HD\,37479 at 687 MHz for one full rotation cycle, we confirm that the star does not produce detectable ECME. This is the only confirmed non-MRP. The confirmed `non-MRPs' are as significant as the confirmed MRPs in order to understand what enables a hot magnetic star to produce ECME.
     \item ECME as a potential magnetospheric probe: We find a double-peaked ECME pulse-profile for the star HD\,182180, which has an obliquity close to $90^\circ$. This provides strong support to the proposition that such pulse-profiles are products of large obliquities \citep{das2022}. In other words, it suggests that ECME properties can provide us with information regarding the stellar magnetospheric parameters. 
    \item The $X$-factor of MRPs: Our data suggest that both incoherent and coherent radio emission produced by hot magnetic stars are powered by the same physical phenomenon (CBO). As a result, the physical quantities that determine the coherent (ECME) radio luminosity are the same as those responsible for the incoherent radio luminosity except for stars with $T_\mathrm{eff}\gtrsim 19$ kK. Thus for the late-B and A-type stars, the coherent radio luminosity is expected to be indifferent to the stellar temperature (or, the mass-loss rate). For the early B (and O-type) stars however, an increasing temperature seems to result in a decreasing coherent radio luminosity. This could be due to the increasing difficulty of satisfying the necessary condition for ECME, which is that the plasma frequency should be smaller than the gyrofrequency. If this explanation is correct, a stronger magnetic field should be able to compensate for the higher temperature. 
    Another possibility is that an increasing temperature results in higher absorption of the emission, in which case, a stronger magnetic field will not be able to balance a higher temperature.
    More observations of early-B type stars will be needed so as to find out which of the two scenarios is correct.
\end{enumerate}

In the future, it will be extremely important to acquire more radio observations of hot magnetic stars. This will allow us to both increase the MRP sample size, as well as to identify confirmed non-MRPs. A special emphasis should be on stars with temperatures above around 19 kK, and stars with longer rotation periods. The latter is important so as to obtain direct evidence of the role of the rotation period in controlling the ECME luminosity.
There are, however, practical challenges in acquiring adequate observations of slow rotators, and one would possibly have to rely on chance detection (if they produce ECME) in all-sky surveys. A sample of MRPs, entirely detected in all-sky surveys (thus not biased by our selection criteria), will provide indirect evidence on whether or not the slow rotators are likely candidates for MRPs. 
The usefulness of such surveys has already been demonstrated for the case of low-mass stars by \citet{callingham2021} where they reported detection of coherent radio emission at 144 MHz from 19 known M-dwarfs using the LOw Frequency ARray Two-meter Sky Survey \citep[LoTSS,][]{shimwell2019}. This large number of discoveries enabled them to infer that such emission is ubiquitous among the M-dwarfs, and also led to the discovery that the radio luminosity is independent of known coronal and chromospheric activity indicators. A similar survey of chemically peculiar stars (spectral types ranging from B8 to A7) using the LoTSS survey, however, turned out to be less fruitful with a detection percentage of only 3\% \citep{hajduk2022}. So far the lowest frequency at which coherent radio emission from a hot magnetic star was observed is 200 MHz \citep[all sky circular polarization survey with the Murchison Widefield array,][]{lenc2018}. This, combined with the fact that no MRP has been confirmed above $\sim 5$ GHz \citep{das2022b} suggest that the suitable frequency range to look for coherent radio emission from MRPs is $\gtrsim 200$ MHz to $\lesssim 5$ GHz.
The ASKAP Variables and Slow Transients (VAST) survey \citep[VAST survey,][]{murphy2013}, the Rapid ASKAP Continuum Survey \citep[RACS,][]{mcconnell2020}, the VLA Sky Survey \citep[VLASS,][]{lacy2016}, the ThunderKAT survey by the MeerKAT \citep{fender2016}, and the all-sky surveys by the upcoming Square Kilometre Array are likely to play important role in enabling us to overcome the current limitations.


\section*{Acknowledgements} 
We thank the anonymous referee for their highly constructive criticism of our work that helped us to improve the manuscript significantly.
We acknowledge support of the Department of Atomic Energy, Government of India, under project no. 12-R\&D-TFR-5.02-0700.
BD acknowledges support from the Bartol Research Institute. M.E.S. acknowledges support from the Annie Jump
Cannon Fellowship, supported by the University of Delaware and endowed by the Mount Cuba Astronomical Observatory. VP acknowledges support by the National Science Foundation under Grant No. AST–1747658. GAW acknowledges Discovery Grant support from the Natural Sciences and Engineering Research Council (NSERC) of Canada. We thank the staff of the GMRT that made these observations possible. 
The GMRT is run by the National Centre for Radio Astrophysics of the Tata Institute 
of Fundamental Research. This research has made use of NASA's Astrophysics Data System.

\section*{Data availability}
The data used in this work are available upon request.



\bibliographystyle{mnras}
\bibliography{das} 

\begin{thebibliography}{}
\makeatletter
\relax
\def\mn@urlcharsother{\let\do\@makeother \do\$\do\&\do\#\do\^\do\_\do\%\do\~}
\def\mn@doi{\begingroup\mn@urlcharsother \@ifnextchar [ {\mn@doi@}
  {\mn@doi@[]}}
\def\mn@doi@[#1]#2{\def\@tempa{#1}\ifx\@tempa\@empty \href
  {http://dx.doi.org/#2} {doi:#2}\else \href {http://dx.doi.org/#2} {#1}\fi
  \endgroup}
\def\mn@eprint#1#2{\mn@eprint@#1:#2::\@nil}
\def\mn@eprint@arXiv#1{\href {http://arxiv.org/abs/#1} {{\tt arXiv:#1}}}
\def\mn@eprint@dblp#1{\href {http://dblp.uni-trier.de/rec/bibtex/#1.xml}
  {dblp:#1}}
\def\mn@eprint@#1:#2:#3:#4\@nil{\def\@tempa {#1}\def\@tempb {#2}\def\@tempc
  {#3}\ifx \@tempc \@empty \let \@tempc \@tempb \let \@tempb \@tempa \fi \ifx
  \@tempb \@empty \def\@tempb {arXiv}\fi \@ifundefined
  {mn@eprint@\@tempb}{\@tempb:\@tempc}{\expandafter \expandafter \csname
  mn@eprint@\@tempb\endcsname \expandafter{\@tempc}}}

\bibitem[\protect\citeauthoryear{{Abt} \& {Levato}}{{Abt} \&
  {Levato}}{1977}]{abt1977}
{Abt} H.~A.,  {Levato} H.,  1977, \mn@doi [\pasp] {10.1086/130230}, \href
  {https://ui.adsabs.harvard.edu/abs/1977PASP...89..797A} {89, 797}

\bibitem[\protect\citeauthoryear{{Bailey} et~al.,}{{Bailey}
  et~al.}{2012}]{bailey2012}
{Bailey} J.~D.,  et~al., 2012, \mn@doi [\mnras]
  {10.1111/j.1365-2966.2012.20881.x}, \href
  {https://ui.adsabs.harvard.edu/abs/2012MNRAS.423..328B} {423, 328}

\bibitem[\protect\citeauthoryear{{Berdyugina}, {Harrington}, {Kuzmychov},
  {Kuhn}, {Hallinan}, {Kowalski}  \& {Hawley}}{{Berdyugina}
  et~al.}{2017}]{berdyugina2017}
{Berdyugina} S.~V.,  {Harrington} D.~M.,  {Kuzmychov} O.,  {Kuhn} J.~R.,
  {Hallinan} G.,  {Kowalski} A.~F.,   {Hawley} S.~L.,  2017, \mn@doi [\apj]
  {10.3847/1538-4357/aa866b}, \href
  {https://ui.adsabs.harvard.edu/abs/2017ApJ...847...61B} {847, 61}

\bibitem[\protect\citeauthoryear{{Callingham} et~al.,}{{Callingham}
  et~al.}{2021}]{callingham2021}
{Callingham} J.~R.,  et~al., 2021, \mn@doi [Nature Astronomy]
  {10.1038/s41550-021-01483-0}, \href
  {https://ui.adsabs.harvard.edu/abs/2021NatAs...5.1233C} {5, 1233}

\bibitem[\protect\citeauthoryear{{Chandra} et~al.,}{{Chandra}
  et~al.}{2015}]{chandra2015}
{Chandra} P.,  et~al., 2015, \mn@doi [\mnras] {10.1093/mnras/stv1378}, \href
  {https://ui.adsabs.harvard.edu/abs/2015MNRAS.452.1245C} {452, 1245}

\bibitem[\protect\citeauthoryear{{Das} \& {Chandra}}{{Das} \&
  {Chandra}}{2021}]{das2021}
{Das} B.,  {Chandra} P.,  2021, \mn@doi [\apj] {10.3847/1538-4357/ac1075},
  \href {https://ui.adsabs.harvard.edu/abs/2021ApJ...921....9D} {921, 9}

\bibitem[\protect\citeauthoryear{{Das}, {Chandra}  \& {Wade}}{{Das}
  et~al.}{2018}]{das2018}
{Das} B.,  {Chandra} P.,   {Wade} G.~A.,  2018, \mn@doi [\mnras]
  {10.1093/mnrasl/slx193}, \href
  {https://ui.adsabs.harvard.edu/abs/2018MNRAS.474L..61D} {474, L61}

\bibitem[\protect\citeauthoryear{{Das}, {Chandra}, {Shultz}  \& {Wade}}{{Das}
  et~al.}{2019a}]{das2019b}
{Das} B.,  {Chandra} P.,  {Shultz} M.~E.,   {Wade} G.~A.,  2019a, \mn@doi
  [\mnras] {10.1093/mnrasl/slz137}, \href
  {https://ui.adsabs.harvard.edu/abs/2019MNRAS.489L.102D} {489, L102}

\bibitem[\protect\citeauthoryear{{Das}, {Chandra}, {Shultz}  \& {Wade}}{{Das}
  et~al.}{2019b}]{das2019a}
{Das} B.,  {Chandra} P.,  {Shultz} M.~E.,   {Wade} G.~A.,  2019b, \mn@doi
  [\apj] {10.3847/1538-4357/ab1b12}, \href
  {https://ui.adsabs.harvard.edu/abs/2019ApJ...877..123D} {877, 123}

\bibitem[\protect\citeauthoryear{{Das}, {Chandra}  \& {Wade}}{{Das}
  et~al.}{2020a}]{das2020b}
{Das} B.,  {Chandra} P.,   {Wade} G.~A.,  2020a, \mn@doi [\mnras]
  {10.1093/mnras/staa2499}, \href
  {https://ui.adsabs.harvard.edu/abs/2020MNRAS.499..702D} {499, 702}

\bibitem[\protect\citeauthoryear{{Das}, {Mondal}  \& {Chandra}}{{Das}
  et~al.}{2020b}]{das2020a}
{Das} B.,  {Mondal} S.,   {Chandra} P.,  2020b, \mn@doi [\apj]
  {10.3847/1538-4357/aba8fd}, \href
  {https://ui.adsabs.harvard.edu/abs/2020ApJ...900..156D} {900, 156}

\bibitem[\protect\citeauthoryear{{Das}, {Chandra}  \& {Petit}}{{Das}
  et~al.}{2022a}]{das2022b}
{Das} B.,  {Chandra} P.,   {Petit} V.,  2022a, \mn@doi [\mnras]
  {10.1093/mnras/stac1894}, \href
  {https://ui.adsabs.harvard.edu/abs/2022MNRAS.515.2008D} {515, 2008}

\bibitem[\protect\citeauthoryear{{Das} et~al.,}{{Das} et~al.}{2022b}]{das2022}
{Das} B.,  et~al., 2022b, \mn@doi [\apj] {10.3847/1538-4357/ac2576}, \href
  {https://ui.adsabs.harvard.edu/abs/2022ApJ...925..125D} {925, 125}

\bibitem[\protect\citeauthoryear{{Davis}, {Vedantham}, {Callingham},
  {Shimwell}, {Vidotto}, {Zarka}, {Ray}  \& {Drabent}}{{Davis}
  et~al.}{2021}]{davis2021}
{Davis} I.,  {Vedantham} H.~K.,  {Callingham} J.~R.,  {Shimwell} T.~W.,
  {Vidotto} A.~A.,  {Zarka} P.,  {Ray} T.~P.,   {Drabent} A.,  2021, arXiv
  e-prints, \href {https://ui.adsabs.harvard.edu/abs/2021arXiv210501021D} {p.
  arXiv:2105.01021}

\bibitem[\protect\citeauthoryear{{Drake}, {Abbott}, {Bastian}, {Bieging},
  {Churchwell}, {Dulk}  \& {Linsky}}{{Drake} et~al.}{1987}]{drake1987}
{Drake} S.~A.,  {Abbott} D.~C.,  {Bastian} T.~S.,  {Bieging} J.~H.,
  {Churchwell} E.,  {Dulk} G.,   {Linsky} J.~L.,  1987, \mn@doi [\apj]
  {10.1086/165784}, \href
  {https://ui.adsabs.harvard.edu/abs/1987ApJ...322..902D} {322, 902}

\bibitem[\protect\citeauthoryear{{Drake}, {Wade}  \& {Linsky}}{{Drake}
  et~al.}{2006}]{drake2006}
{Drake} S.~A.,  {Wade} G.~A.,   {Linsky} J.~L.,  2006, in {Wilson} A.,  ed.,
  ESA Special Publication Vol. 604, The X-ray Universe 2005. p.~73

\bibitem[\protect\citeauthoryear{{Eikenberry} et~al.,}{{Eikenberry}
  et~al.}{2014}]{eikenberry2014}
{Eikenberry} S.~S.,  et~al., 2014, \mn@doi [\apjl]
  {10.1088/2041-8205/784/2/L30}, \href
  {https://ui.adsabs.harvard.edu/abs/2014ApJ...784L..30E} {784, L30}

\bibitem[\protect\citeauthoryear{{Fender} et~al.,}{{Fender}
  et~al.}{2016}]{fender2016}
{Fender} R.,  et~al., 2016, in MeerKAT Science: On the Pathway to the SKA.
  p.~13 (\mn@eprint {arXiv} {1711.04132})

\bibitem[\protect\citeauthoryear{{Gaia Collaboration} et~al.,}{{Gaia
  Collaboration} et~al.}{2018}]{gaia2018}
{Gaia Collaboration} et~al., 2018, \mn@doi [\aap]
  {10.1051/0004-6361/201833051}, \href
  {https://ui.adsabs.harvard.edu/abs/2018A&A...616A...1G} {616, A1}

\bibitem[\protect\citeauthoryear{{Gupta} et~al.,}{{Gupta}
  et~al.}{2017}]{gupta2017}
{Gupta} Y.,  et~al., 2017, \mn@doi [Current Science]
  {10.18520/cs/v113/i04/707-714}, \href
  {https://ui.adsabs.harvard.edu/abs/2017CSci..113..707G} {113, 707}

\bibitem[\protect\citeauthoryear{{Hajduk}, {Leto}, {Vedantham}, {Trigilio},
  {Haverkorn}, {Shimwell}, {Callingham}  \& {White}}{{Hajduk}
  et~al.}{2022}]{hajduk2022}
{Hajduk} M.,  {Leto} P.,  {Vedantham} H.,  {Trigilio} C.,  {Haverkorn} M.,
  {Shimwell} T.,  {Callingham} J.~R.,   {White} G.~J.,  2022, \mn@doi [A\&A]
  {10.1051/0004-6361/202243784}, 665, A152

\bibitem[\protect\citeauthoryear{{Hallinan} et~al.,}{{Hallinan}
  et~al.}{2007}]{hallinan2007}
{Hallinan} G.,  et~al., 2007, \mn@doi [\apjl] {10.1086/519790}, \href
  {https://ui.adsabs.harvard.edu/abs/2007ApJ...663L..25H} {663, L25}

\bibitem[\protect\citeauthoryear{{Houk}}{{Houk}}{1982}]{houk1982}
{Houk} N.,  1982, {Michigan Catalogue of Two-dimensional Spectral Types for the
  HD stars. Volume\_3. Declinations -40\_{\textflorin}0 to
  -26\_{\textflorin}0.}

\bibitem[\protect\citeauthoryear{{Houk} \& {Swift}}{{Houk} \&
  {Swift}}{1999}]{houk1999}
{Houk} N.,  {Swift} C.,  1999, Michigan Spectral Survey, \href
  {https://ui.adsabs.harvard.edu/abs/1999MSS...C05....0H} {5, 0}

\bibitem[\protect\citeauthoryear{{Hunger}, {Heber}  \& {Groote}}{{Hunger}
  et~al.}{1989}]{hunger1989}
{Hunger} K.,  {Heber} U.,   {Groote} D.,  1989, \aap, \href
  {https://ui.adsabs.harvard.edu/abs/1989A&A...224...57H} {224, 57}

\bibitem[\protect\citeauthoryear{{Kao}, {Hallinan}, {Pineda}, {Stevenson}  \&
  {Burgasser}}{{Kao} et~al.}{2018}]{kao2018}
{Kao} M.~M.,  {Hallinan} G.,  {Pineda} J.~S.,  {Stevenson} D.,   {Burgasser}
  A.,  2018, \mn@doi [\apjs] {10.3847/1538-4365/aac2d5}, \href
  {https://ui.adsabs.harvard.edu/abs/2018ApJS..237...25K} {237, 25}

\bibitem[\protect\citeauthoryear{{Kavanagh}, {Vidotto}, {Vedantham}, {Jardine},
  {Callingham}  \& {Morin}}{{Kavanagh} et~al.}{2022}]{kavanagh2022}
{Kavanagh} R.~D.,  {Vidotto} A.~A.,  {Vedantham} H.~K.,  {Jardine} M.~M.,
  {Callingham} J.~R.,   {Morin} J.,  2022, \mn@doi [\mnras]
  {10.1093/mnras/stac1264}, \href
  {https://ui.adsabs.harvard.edu/abs/2022MNRAS.514..675K} {514, 675}

\bibitem[\protect\citeauthoryear{{Kochukhov}}{{Kochukhov}}{2021}]{kochukhov2021}
{Kochukhov} O.,  2021, \mn@doi [\aapr] {10.1007/s00159-020-00130-3}, \href
  {https://ui.adsabs.harvard.edu/abs/2021A&ARv..29....1K} {29, 1}

\bibitem[\protect\citeauthoryear{{Kochukhov}, {L{\"u}ftinger}, {Neiner},
  {Alecian}  \& {MiMeS Collaboration}}{{Kochukhov}
  et~al.}{2014}]{kochukhov2014}
{Kochukhov} O.,  {L{\"u}ftinger} T.,  {Neiner} C.,  {Alecian} E.,   {MiMeS
  Collaboration} 2014, \mn@doi [\aap] {10.1051/0004-6361/201423472}, \href
  {https://ui.adsabs.harvard.edu/abs/2014A&A...565A..83K} {565, A83}

\bibitem[\protect\citeauthoryear{{Kochukhov}, {Silvester}, {Bailey}, {Land
  street}  \& {Wade}}{{Kochukhov} et~al.}{2017}]{kochukhov2017}
{Kochukhov} O.,  {Silvester} J.,  {Bailey} J.~D.,  {Land street} J.~D.,
  {Wade} G.~A.,  2017, \mn@doi [\aap] {10.1051/0004-6361/201730919}, \href
  {https://ui.adsabs.harvard.edu/abs/2017A&A...605A..13K} {605, A13}

\bibitem[\protect\citeauthoryear{{Kochukhov}, {Shultz}  \&
  {Neiner}}{{Kochukhov} et~al.}{2019}]{kochukhov2019}
{Kochukhov} O.,  {Shultz} M.,   {Neiner} C.,  2019, \mn@doi [\aap]
  {10.1051/0004-6361/201834279}, \href
  {https://ui.adsabs.harvard.edu/abs/2019A&A...621A..47K} {621, A47}

\bibitem[\protect\citeauthoryear{{Lacy}, {Baum}, {Chandler}, {Chatterjee},
  {Murphy}, {Myers}  \& {VLASS Survey Science Group}}{{Lacy}
  et~al.}{2016}]{lacy2016}
{Lacy} M.,  {Baum} S.~A.,  {Chandler} C.~J.,  {Chatterjee} S.,  {Murphy} E.~J.,
   {Myers} S.~T.,   {VLASS Survey Science Group} 2016, in American Astronomical
  Society Meeting Abstracts \#227. p. 324.09

\bibitem[\protect\citeauthoryear{{Landstreet} \& {Borra}}{{Landstreet} \&
  {Borra}}{1978}]{landstreet1978}
{Landstreet} J.~D.,  {Borra} E.~F.,  1978, \mn@doi [\apjl] {10.1086/182746},
  \href {https://ui.adsabs.harvard.edu/abs/1978ApJ...224L...5L} {224, L5}

\bibitem[\protect\citeauthoryear{{Landstreet} \& {Mathys}}{{Landstreet} \&
  {Mathys}}{2000}]{landstreet2000}
{Landstreet} J.~D.,  {Mathys} G.,  2000, \aap, \href
  {https://ui.adsabs.harvard.edu/abs/2000A&A...359..213L} {359, 213}

\bibitem[\protect\citeauthoryear{{Lenc}, {Murphy}, {Lynch}, {Kaplan}  \&
  {Zhang}}{{Lenc} et~al.}{2018}]{lenc2018}
{Lenc} E.,  {Murphy} T.,  {Lynch} C.~R.,  {Kaplan} D.~L.,   {Zhang} S.~N.,
  2018, \mn@doi [\mnras] {10.1093/mnras/sty1304}, \href
  {https://ui.adsabs.harvard.edu/abs/2018MNRAS.478.2835L} {478, 2835}

\bibitem[\protect\citeauthoryear{{Leone}, {Umana}  \& {Trigilio}}{{Leone}
  et~al.}{1996}]{leone1996}
{Leone} F.,  {Umana} G.,   {Trigilio} C.,  1996, \aap, \href
  {https://ui.adsabs.harvard.edu/abs/1996A&A...310..271L} {310, 271}

\bibitem[\protect\citeauthoryear{{Leone}, {Trigilio}, {Neri}  \&
  {Umana}}{{Leone} et~al.}{2004}]{leone2004}
{Leone} F.,  {Trigilio} C.,  {Neri} R.,   {Umana} G.,  2004, \mn@doi [\aap]
  {10.1051/0004-6361:20040181}, \href
  {https://ui.adsabs.harvard.edu/abs/2004A&A...423.1095L} {423, 1095}

\bibitem[\protect\citeauthoryear{{Leto}, {Trigilio}, {Buemi}, {Leone}  \&
  {Umana}}{{Leto} et~al.}{2012}]{leto2012}
{Leto} P.,  {Trigilio} C.,  {Buemi} C.~S.,  {Leone} F.,   {Umana} G.,  2012,
  \mn@doi [\mnras] {10.1111/j.1365-2966.2012.20997.x}, \href
  {https://ui.adsabs.harvard.edu/abs/2012MNRAS.423.1766L} {423, 1766}

\bibitem[\protect\citeauthoryear{{Leto} et~al.,}{{Leto}
  et~al.}{2017}]{leto2017}
{Leto} P.,  et~al., 2017, \mn@doi [\mnras] {10.1093/mnras/stx267}, \href
  {https://ui.adsabs.harvard.edu/abs/2017MNRAS.467.2820L} {467, 2820}

\bibitem[\protect\citeauthoryear{{Leto} et~al.,}{{Leto}
  et~al.}{2019}]{leto2019}
{Leto} P.,  et~al., 2019, \mn@doi [\mnras] {10.1093/mnrasl/sly179}, \href
  {https://ui.adsabs.harvard.edu/abs/2019MNRAS.482L...4L} {482, L4}

\bibitem[\protect\citeauthoryear{{Leto} et~al.,}{{Leto}
  et~al.}{2020a}]{leto2020}
{Leto} P.,  et~al., 2020a, \mn@doi [\mnras] {10.1093/mnras/staa587}, \href
  {https://ui.adsabs.harvard.edu/abs/2020MNRAS.493.4657L} {493, 4657}

\bibitem[\protect\citeauthoryear{{Leto} et~al.,}{{Leto}
  et~al.}{2020b}]{leto2020b}
{Leto} P.,  et~al., 2020b, \mn@doi [\mnras] {10.1093/mnrasl/slaa157}, \href
  {https://ui.adsabs.harvard.edu/abs/2020MNRAS.499L..72L} {499, L72}

\bibitem[\protect\citeauthoryear{{Leto} et~al.,}{{Leto}
  et~al.}{2021}]{leto2021}
{Leto} P.,  et~al., 2021, \mn@doi [\mnras] {10.1093/mnras/stab2168}, \href
  {https://ui.adsabs.harvard.edu/abs/2021MNRAS.507.1979L} {507, 1979}

\bibitem[\protect\citeauthoryear{{Lim}, {Drake}  \& {Linsky}}{{Lim}
  et~al.}{1996}]{lim1996}
{Lim} J.,  {Drake} S.~A.,   {Linsky} J.~L.,  1996, {Rotational Modulation of
  Radio Emission from the Magnetic BP Star HR 5624}.
p.~324

\bibitem[\protect\citeauthoryear{{Linsky}, {Drake}  \& {Bastian}}{{Linsky}
  et~al.}{1992}]{linsky1992}
{Linsky} J.~L.,  {Drake} S.~A.,   {Bastian} T.~S.,  1992, \mn@doi [\apj]
  {10.1086/171509}, \href
  {https://ui.adsabs.harvard.edu/abs/1992ApJ...393..341L} {393, 341}

\bibitem[\protect\citeauthoryear{{Llama}, {Jardine}, {Wood}, {Hallinan}  \&
  {Morin}}{{Llama} et~al.}{2018}]{llama2018}
{Llama} J.,  {Jardine} M.~M.,  {Wood} K.,  {Hallinan} G.,   {Morin} J.,  2018,
  \mn@doi [\apj] {10.3847/1538-4357/aaa59f}, \href
  {https://ui.adsabs.harvard.edu/abs/2018ApJ...854....7L} {854, 7}

\bibitem[\protect\citeauthoryear{{McConnell} et~al.,}{{McConnell}
  et~al.}{2020}]{mcconnell2020}
{McConnell} D.,  et~al., 2020, \mn@doi [\pasa] {10.1017/pasa.2020.41}, \href
  {https://ui.adsabs.harvard.edu/abs/2020PASA...37...48M} {37, e048}

\bibitem[\protect\citeauthoryear{{McMullin}, {Waters}, {Schiebel}, {Young}  \&
  {Golap}}{{McMullin} et~al.}{2007}]{mcmullin2007}
{McMullin} J.~P.,  {Waters} B.,  {Schiebel} D.,  {Young} W.,   {Golap} K.,
  2007, in {Shaw} R.~A.,  {Hill} F.,   {Bell} D.~J.,  eds,  Astronomical
  Society of the Pacific Conference Series Vol. 376, Astronomical Data Analysis
  Software and Systems XVI. p.~127

\bibitem[\protect\citeauthoryear{{Melrose} \& {Dulk}}{{Melrose} \&
  {Dulk}}{1982}]{melrose1982}
{Melrose} D.~B.,  {Dulk} G.~A.,  1982, \mn@doi [\apj] {10.1086/160219}, \href
  {https://ui.adsabs.harvard.edu/abs/1982ApJ...259..844M} {259, 844}

\bibitem[\protect\citeauthoryear{{Mikul{\'a}{\v{s}}ek}}{{Mikul{\'a}{\v{s}}ek}}{2016}]{mikulasek2016}
{Mikul{\'a}{\v{s}}ek} Z.,  2016, Contributions of the Astronomical Observatory
  Skalnate Pleso, \href {https://ui.adsabs.harvard.edu/abs/2016CoSka..46...95M}
  {46, 95}

\bibitem[\protect\citeauthoryear{{Murphy} et~al.,}{{Murphy}
  et~al.}{2013}]{murphy2013}
{Murphy} T.,  et~al., 2013, \mn@doi [\pasa] {10.1017/pasa.2012.006}, \href
  {https://ui.adsabs.harvard.edu/abs/2013PASA...30....6M} {30, e006}

\bibitem[\protect\citeauthoryear{{Naz{\'e}}, {Petit}, {Rinbrand}, {Cohen},
  {Owocki}, {ud-Doula}  \& {Wade}}{{Naz{\'e}} et~al.}{2014}]{naze2014}
{Naz{\'e}} Y.,  {Petit} V.,  {Rinbrand} M.,  {Cohen} D.,  {Owocki} S.,
  {ud-Doula} A.,   {Wade} G.~A.,  2014, \mn@doi [\apjs]
  {10.1088/0067-0049/215/1/10}, \href
  {https://ui.adsabs.harvard.edu/abs/2014ApJS..215...10N} {215, 10}

\bibitem[\protect\citeauthoryear{{Oksala}, {Wade}, {Marcolino}, {Grunhut},
  {Bohlender}, {Manset}, {Townsend}  \& {Mimes Collaboration}}{{Oksala}
  et~al.}{2010}]{oksala2010}
{Oksala} M.~E.,  {Wade} G.~A.,  {Marcolino} W.~L.~F.,  {Grunhut} J.,
  {Bohlender} D.,  {Manset} N.,  {Townsend} R.~H.~D.,   {Mimes Collaboration}
  2010, \mn@doi [\mnras] {10.1111/j.1745-3933.2010.00857.x}, \href
  {https://ui.adsabs.harvard.edu/abs/2010MNRAS.405L..51O} {405, L51}

\bibitem[\protect\citeauthoryear{{Oksala}, {Wade}, {Townsend}, {Owocki},
  {Kochukhov}, {Neiner}, {Alecian}  \& {Grunhut}}{{Oksala}
  et~al.}{2012}]{oksala2012}
{Oksala} M.~E.,  {Wade} G.~A.,  {Townsend} R.~H.~D.,  {Owocki} S.~P.,
  {Kochukhov} O.,  {Neiner} C.,  {Alecian} E.,   {Grunhut} J.,  2012, \mn@doi
  [\mnras] {10.1111/j.1365-2966.2011.19753.x}, \href
  {https://ui.adsabs.harvard.edu/abs/2012MNRAS.419..959O} {419, 959}

\bibitem[\protect\citeauthoryear{{Oksala} et~al.,}{{Oksala}
  et~al.}{2015}]{oksala2015}
{Oksala} M.~E.,  et~al., 2015, \mn@doi [\mnras] {10.1093/mnras/stv1086}, \href
  {https://ui.adsabs.harvard.edu/abs/2015MNRAS.451.2015O} {451, 2015}

\bibitem[\protect\citeauthoryear{{Owocki}, {Shultz}, {ud-Doula}, {Sundqvist},
  {Townsend}  \& {Cranmer}}{{Owocki} et~al.}{2020}]{owocki2020}
{Owocki} S.~P.,  {Shultz} M.~E.,  {ud-Doula} A.,  {Sundqvist} J.~O.,
  {Townsend} R. H.~D.,   {Cranmer} S.~R.,  2020, \mn@doi [\mnras]
  {10.1093/mnras/staa2325}, \href
  {https://ui.adsabs.harvard.edu/abs/2020MNRAS.499.5366O} {499, 5366}

\bibitem[\protect\citeauthoryear{{Owocki}, {Shultz}, {ud-Doula}, {Chandra},
  {Das}  \& {Leto}}{{Owocki} et~al.}{2022}]{owocki2022}
{Owocki} S.~P.,  {Shultz} M.~E.,  {ud-Doula} A.,  {Chandra} P.,  {Das} B.,
  {Leto} P.,  2022, \mn@doi [\mnras] {10.1093/mnras/stac341}, \href
  {https://ui.adsabs.harvard.edu/abs/2022MNRAS.tmp.1106O} {}

\bibitem[\protect\citeauthoryear{{P{\'e}rez-Torres} et~al.,}{{P{\'e}rez-Torres}
  et~al.}{2021}]{perez-torres2021}
{P{\'e}rez-Torres} M.,  et~al., 2021, \mn@doi [\aap]
  {10.1051/0004-6361/202039052}, \href
  {https://ui.adsabs.harvard.edu/abs/2021A&A...645A..77P} {645, A77}

\bibitem[\protect\citeauthoryear{{Petit} et~al.,}{{Petit}
  et~al.}{2013}]{petit2013}
{Petit} V.,  et~al., 2013, \mn@doi [\mnras] {10.1093/mnras/sts344}, \href
  {https://ui.adsabs.harvard.edu/abs/2013MNRAS.429..398P} {429, 398}

\bibitem[\protect\citeauthoryear{{Pritchard} et~al.,}{{Pritchard}
  et~al.}{2021}]{pritchard2021}
{Pritchard} J.,  et~al., 2021, \mn@doi [\mnras] {10.1093/mnras/stab299}, \href
  {https://ui.adsabs.harvard.edu/abs/2021MNRAS.502.5438P} {502, 5438}

\bibitem[\protect\citeauthoryear{{Rivinius}, {Szeifert}, {Barrera}, {Townsend},
  {{\v{S}}tefl}  \& {Baade}}{{Rivinius} et~al.}{2010}]{rivinius2010}
{Rivinius} T.,  {Szeifert} T.,  {Barrera} L.,  {Townsend} R.~H.~D.,
  {{\v{S}}tefl} S.,   {Baade} D.,  2010, \mn@doi [\mnras]
  {10.1111/j.1745-3933.2010.00856.x}, \href
  {https://ui.adsabs.harvard.edu/abs/2010MNRAS.405L..46R} {405, L46}

\bibitem[\protect\citeauthoryear{{Rivinius}, {Townsend}, {Kochukhov},
  {{\v{S}}tefl}, {Baade}, {Barrera}  \& {Szeifert}}{{Rivinius}
  et~al.}{2013}]{rivinius2013}
{Rivinius} T.,  {Townsend} R.~H.~D.,  {Kochukhov} O.,  {{\v{S}}tefl} S.,
  {Baade} D.,  {Barrera} L.,   {Szeifert} T.,  2013, \mn@doi [\mnras]
  {10.1093/mnras/sts323}, \href
  {https://ui.adsabs.harvard.edu/abs/2013MNRAS.429..177R} {429, 177}

\bibitem[\protect\citeauthoryear{{Semel}}{{Semel}}{1989}]{semel1989}
{Semel} M.,  1989, \aap, \href
  {https://ui.adsabs.harvard.edu/abs/1989A&A...225..456S} {225, 456}

\bibitem[\protect\citeauthoryear{{Shimwell} et~al.,}{{Shimwell}
  et~al.}{2019}]{shimwell2019}
{Shimwell} T.~W.,  et~al., 2019, \mn@doi [\aap] {10.1051/0004-6361/201833559},
  \href {https://ui.adsabs.harvard.edu/abs/2019A&A...622A...1S} {622, A1}

\bibitem[\protect\citeauthoryear{{Shultz} et~al.,}{{Shultz}
  et~al.}{2015}]{shultz2015}
{Shultz} M.,  et~al., 2015, \mn@doi [\mnras] {10.1093/mnras/stv564}, \href
  {https://ui.adsabs.harvard.edu/abs/2015MNRAS.449.3945S} {449, 3945}

\bibitem[\protect\citeauthoryear{{Shultz} et~al.,}{{Shultz}
  et~al.}{2018}]{shultz2018}
{Shultz} M.~E.,  et~al., 2018, \mn@doi [\mnras] {10.1093/mnras/sty103}, \href
  {https://ui.adsabs.harvard.edu/abs/2018MNRAS.475.5144S} {475, 5144}

\bibitem[\protect\citeauthoryear{{Shultz} et~al.,}{{Shultz}
  et~al.}{2019a}]{shultz2019b}
{Shultz} M.~E.,  et~al., 2019a, \mn@doi [\mnras] {10.1093/mnras/stz416}, \href
  {https://ui.adsabs.harvard.edu/abs/2019MNRAS.485.1508S} {485, 1508}

\bibitem[\protect\citeauthoryear{{Shultz} et~al.,}{{Shultz}
  et~al.}{2019b}]{shultz2019c}
{Shultz} M.~E.,  et~al., 2019b, \mn@doi [\mnras] {10.1093/mnras/stz2551}, \href
  {https://ui.adsabs.harvard.edu/abs/2019MNRAS.490..274S} {490, 274}

\bibitem[\protect\citeauthoryear{{Shultz} et~al.,}{{Shultz}
  et~al.}{2020}]{shultz2020}
{Shultz} M.~E.,  et~al., 2020, \mn@doi [\mnras] {10.1093/mnras/staa3102}, \href
  {https://ui.adsabs.harvard.edu/abs/2020MNRAS.499.5379S} {499, 5379}

\bibitem[\protect\citeauthoryear{{Shultz} et~al.,}{{Shultz}
  et~al.}{2022}]{shultz2022}
{Shultz} M.~E.,  et~al., 2022, \mn@doi [\mnras] {10.1093/mnras/stac136}, \href
  {https://ui.adsabs.harvard.edu/abs/2022MNRAS.tmp.1099S} {}

\bibitem[\protect\citeauthoryear{{Sikora} et~al.,}{{Sikora}
  et~al.}{2016}]{sikora2016}
{Sikora} J.,  et~al., 2016, \mn@doi [\mnras] {10.1093/mnras/stw1077}, \href
  {https://ui.adsabs.harvard.edu/abs/2016MNRAS.460.1811S} {460, 1811}

\bibitem[\protect\citeauthoryear{{Swarup}, {Ananthakrishnan}, {Kapahi}, {Rao},
  {Subrahmanya}  \& {Kulkarni}}{{Swarup} et~al.}{1991}]{swarup1991}
{Swarup} G.,  {Ananthakrishnan} S.,  {Kapahi} V.~K.,  {Rao} A.~P.,
  {Subrahmanya} C.~R.,   {Kulkarni} V.~K.,  1991, Current Science, \href
  {https://ui.adsabs.harvard.edu/abs/1991CSci...60...95S} {60, 95}

\bibitem[\protect\citeauthoryear{{Townsend} \& {Owocki}}{{Townsend} \&
  {Owocki}}{2005}]{townsend2005}
{Townsend} R.~H.~D.,  {Owocki} S.~P.,  2005, \mn@doi [\mnras]
  {10.1111/j.1365-2966.2005.08642.x}, \href
  {https://ui.adsabs.harvard.edu/abs/2005MNRAS.357..251T} {357, 251}

\bibitem[\protect\citeauthoryear{{Townsend}, {Oksala}, {Cohen}, {Owocki}  \&
  {ud-Doula}}{{Townsend} et~al.}{2010}]{townsend2010}
{Townsend} R.~H.~D.,  {Oksala} M.~E.,  {Cohen} D.~H.,  {Owocki} S.~P.,
  {ud-Doula} A.,  2010, \mn@doi [\apjl] {10.1088/2041-8205/714/2/L318}, \href
  {https://ui.adsabs.harvard.edu/abs/2010ApJ...714L.318T} {714, L318}

\bibitem[\protect\citeauthoryear{{Trigilio}, {Leto}, {Leone}, {Umana}  \&
  {Buemi}}{{Trigilio} et~al.}{2000}]{trigilio2000}
{Trigilio} C.,  {Leto} P.,  {Leone} F.,  {Umana} G.,   {Buemi} C.,  2000, \aap,
  \href {https://ui.adsabs.harvard.edu/abs/2000A&A...362..281T} {362, 281}

\bibitem[\protect\citeauthoryear{{Trigilio}, {Leto}, {Umana}, {Buemi}  \&
  {Leone}}{{Trigilio} et~al.}{2011}]{trigilio2011}
{Trigilio} C.,  {Leto} P.,  {Umana} G.,  {Buemi} C.~S.,   {Leone} F.,  2011,
  \mn@doi [\apjl] {10.1088/2041-8205/739/1/L10}, \href
  {https://ui.adsabs.harvard.edu/abs/2011ApJ...739L..10T} {739, L10}

\bibitem[\protect\citeauthoryear{{Ud-Doula}, {Owocki}  \&
  {Townsend}}{{Ud-Doula} et~al.}{2008}]{ud-doula2008}
{Ud-Doula} A.,  {Owocki} S.~P.,   {Townsend} R. H.~D.,  2008, \mn@doi [\mnras]
  {10.1111/j.1365-2966.2008.12840.x}, \href
  {https://ui.adsabs.harvard.edu/abs/2008MNRAS.385...97U} {385, 97}

\bibitem[\protect\citeauthoryear{{Yakunin} et~al.,}{{Yakunin}
  et~al.}{2015}]{yakunin2015}
{Yakunin} I.,  et~al., 2015, \mn@doi [\mnras] {10.1093/mnras/stu2401}, \href
  {https://ui.adsabs.harvard.edu/abs/2015MNRAS.447.1418Y} {447, 1418}

\makeatother
\end{thebibliography}




\bsp	

\label{lastpage}
\end{document}